\documentclass[aip,amssymb,amsfonts,graphicx]{revtex4-1}
\usepackage{graphics}
\usepackage{amsmath}
\usepackage{graphicx}
\usepackage{color}
\usepackage{comment}
\usepackage{lineno}
\begin{document}
\title{Mechanisms for bump state localization in two-dimensional networks of leaky Integrate-and-Fire neurons}

\author{A. Provata}
\affiliation{Institute of Nanoscience and Nanotechnology, National Center for Scientific Research ``Demokritos'', 15341 Athens, Greece}
\author{J. Hizanidis}
\affiliation{Institute of Electronic Structure and Laser, Foundation for Research and Technology-Hellas, 70013 Herakleio, Crete, Greece}
\affiliation{Institute of Nanoscience and Nanotechnology, National Center for Scientific Research ``Demokritos'', 15341 Athens, Greece} 

\author{K. Anesiadis}
\affiliation{Institute of Nanoscience and Nanotechnology, National Center for Scientific Research ``Demokritos'', 15341 Athens, Greece}

\author{O. E. Omel'chenko}
\affiliation{Department of Physics and Astronomy, University of Potsdam, Karl-Liebknecht-Str. 24/25, D-14476, Potsdam-Golm, Germany} 
%
%

%
\date{Received: \today / Revised version: date}
%

\begin{abstract}
\centerline{\bf Abstract}
Networks of nonlocally coupled leaky Integrate-and-Fire neurons exhibit a variety of complex collective behaviors, such as partial synchronization, frequency or amplitude
chimeras,
solitary states and bump states. In particular, the bump states  consist of one 
or many regions of asynchronous elements within a sea of 
subthreshold (quiescent) elements. The asynchronous domains travel in the network in a
direction predetermined by the initial conditions. In the present study we investigate
the occurrence of bump states in networks of leaky Integrate-and-Fire neurons
in two-dimensions using nonlocal toroidal connectivity and we
explore possible mechanisms for stabilizing the moving asynchronous domains.
Our findings indicate that I) incorporating a refractory period can effectively 
anchor the position of these domains in the network, and II) the switching off of 
some randomly preselected nodes (i.e., making them permanently idle/inactive)
 can likewise contribute to
stabilizing the positions of the asynchronous domains. 
In particular, in case II for large values of the coupling strength and a large percentage of idle
elements, all nodes acquire different fixed (frozen) values in 
the quiescent
region and oscillations cease throughout the network due to self-organization.  
For the special case of stationary bump states,
we propose an analytical approach to predict their properties.
This approach is based on the self-consistency argument
and is valid for infinitely large networks.
Case I is of
particular biomedical interest in view of the importance of refractoriness
for biological neurons, while case II can be biomedically relevant when
designing therapeutic methods for stabilizing moving signals in the brain.
\end{abstract}


\maketitle

%

\begin{quotation}
In networks of interacting neurons, bump states consist of clusters of active (firing) neurons co-existing with quiescent 
ones which form a passive background. The active clusters may be spatially localized, moving randomly, or traveling 
in a specific direction in the network. The mechanism behind the formation and motion of these complex spatiotemporal 
states is still under intense investigations. Bump states are considered to be important in brain dynamics, 
notably in brain waves, in the mechanism of visual orientation tuning, in the head direction system, 
and in working memory. Inspired by the neurophysiology of nerve cells, the present study explores possible 
techniques for controlling the motion of bump states, which are capable of spatially confining these states, 
namely, the refractory period and the presence of idle nodes in the network. After firing, the neurons spend a time 
interval, the so-called refractory period, during which they can neither integrate their potential nor produce 
new spikes. This period of neuronal rest encourages unidirectional flow of action potentials because they cannot 
travel backwards to inactive neurons. The presence of idle nodes in the network is also neurophysiologically 
relevant because they may occur due to aging, traumas, or diseases. We numerically show that both of these 
neurophysiological mechanisms can control the motion of bump states and, therefore, can be held responsible for 
self regulating the brain activity.
\end{quotation}

\section{Introduction}
\label{intro}


\par Bump states are encountered in networks of coupled elements and they are characterized by coexistence
of active and quiescent regions which are simultaneously present in the network 
\cite{laing:2001,laing:2020,laing:2021}. 
These are nontrivial states, especially in the cases where
all network nodes have identical dynamics and are identically coupled. Bump states are different from chimera states
in the sense that the latter are characterized by coexistence of two types of active
regions: coherent and incoherent \cite{kuramoto:2002,abrams:2004}.
That is to say, while in chimera states all elements are active with different frequencies (for frequency chimeras) or 
amplitudes (for amplitude chimeras) dominating in the different
regions, in bump states some areas present activity (oscillations) while other areas are quiescent. 
Bump states  have been reported to be relevant in brain dynamics (brain waves)  \cite{ermentrout:1998},
in the mechanism
of visual orientation tuning \cite{somers:1995}, in the head direction system
\cite{redish:1996,poll:2016}, and in working memory 
\cite{ermentrout:1998,wilson:1972,wilson:1973,amit:1997,laing:2002,wimmer:2014}.

In the recent literature, chimeras and bump states have been investigated as hybrid dynamical states/regimes resulting from interactions between multiple elements in complex networks. Chimera states, as prototype hybrid patterns, have attracted 
much attention
after their first discovery in the early 2000s \cite{kuramoto:2002,abrams:2004} and they have been reported for different coupling schemes and types of nonlinear oscillators, such as
 FitzHugh-Nagumo 
\cite{omelchenko:2013,zakharova:2017,schmidt:2017,shepelev:2019}, 
van der Pol \cite{ulonska:2016}, Ginzburg-Landau \cite{sethia:2014},
Stuart-Landau oscillators \cite{yeldesbay:2014,gjurchinovski:2017,tumash:2019},
Integrate-and-Fire neurons \cite{luccioli:2010,olmi:2010,olmi:2015,politi:2018,ullner:2020,schmidt:2017}, and superconducting oscillators \cite{hizanidis:2016,hizanidis:2018},
to name just a few. The network connectivity has also contributed to the rich variety of these hybrid patterns 
with studies in 1D \cite{omelchenko:2013,provata:2020}, 
2D \cite{schmidt:2017,maistrenko:2017,laing:2017,oomelchenko:2019,hizanidis:2020} 
and 3D \cite{maistrenko:2017,oomelchenko:2019,maistrenko:2020,kasimatis:2018,koulierakis:2020} networks, while 
diverse studies have confirmed the existence 
of these states experimentally
\cite{hagerstrom:2012,gambuzza:2014,tinsley:2012,wickramasinghe:2013,schmidt:2014,kapitaniak:2014,dudkowski:2016}.  
Bump states, on the other hand, have been introduced earlier than chimera states 
in the framework of networks of interacting 
neurons simulating working memory effects \cite{wilson:1972,wilson:1973,ermentrout:1998,laing:2001}. 
Most recently they have gained attention on their own, as a mechanism for complex dynamical pattern 
formation,
see \cite{oomelchenko:2021,laing:2020,laing:2021,tsigkri:2017,rontogiannis:2021,provata:2024}.
Similarly to chimera states, bump states are now discovered numerically for diverse dynamical systems,
such as the theta neuron network \cite{laing:2020,laing:2021,oomelchenko:2022b,laing:2023},
the Integrate-and-Fire model \cite{tsigkri:2017},
 the quadratic Integrate-and-Fire neurons \cite{byrne:2019,schmidt:2020,oomelchenko:2024},
the FitzHugh-Nagumo oscillators \cite{rontogiannis:2021}, 
and other models \cite{provata:2024,oomelchenko:2021}.

\par In the numerical simulations of identical and  identically interacting elements, when
 bump states are produced the asynchronous oscillating
regions (also called simply ``bumps'') present a directed motion in the ``sea'' composed 
of quiescent elements. The actual direction of the bump motion
depends on the initial network conditions. There is an open question on the role of moving vs spatially 
localized bump states
in real-world networks and especially in brain networks, where 
neuronal models find applications. This question has partially been addressed, mainly in 1D ring geometries,
 for the cases of Integrate-and-Fire networks with long-distance spatial couplings
 and time delays \cite{laing:2001,laing:2001a}, the theta neuron \cite{laing:2020} 
and the Kuramoto model \cite{laing:2011}. In this view,
we here propose two processes as possible mechanisms for stabilizing the bump motion.
Both proposed mechanisms may find a counterpart in the actual dynamics of the brain.
Mechanism I refers to the introduction of a refractory period. In fact, 
refractoriness is an integral part of the realistic
neuronal dynamics; all neurons are known to undergo a refractory period after spiking. During refractoriness
the neurons remain unresponsive, even if they receive external signals capable of producing spikes. Mechanism II
refers to the presence of specific randomly distributed nodes (neurons) which are 
permanently inactive, also referred to as ``dead'' or idle nodes.
In realistic brain networks dead neurons may appear due to brain illness, local traumas or aging processes.
Both proposed mechanisms, I and II, have neurological references and we will demonstrate in the sequel that
they lead to spatial confinement of the asynchronous oscillating regions under specific conditions, quantifiable
 by the network parameters. 

\par In 1D ring networks of Leaky Integrate-and-Fire (LIF) neurons,
 traveling and localized bump states have been previously reported 
in Ref.~\cite{tsigkri:2017} where, besides nonlocal connectivity, reflecting connectivity was also investigated.
In Ref.~\cite{anesiadis:2022} the numerical studies have demonstrated the presence of bump states in multiplex 
ring networks. Furthermore, the relationship between bump attractors and traveling waves
 in a classical 1D network of excitable, leaky Integrate-and-Fire neurons is studied in Ref.~\cite{avitabile:2023}. 
As stated in the last reference, a rigorous  analysis of these nonsmooth 
networks is still an open and challenging problem.
The simulations presented in the current study concern bump states in 2D.
Two- and three-dimensional patterns are 
less addressed in the scientific literature because of two main reasons: 
a) interesting, unexpected emergent features are already observed in 1D systems 
and scientific efforts tend to concentrate on these, 
before addressing higher dimensional systems and b)
2D (and 3D) simulations are computationally expensive, require long 
computational times and considerable computational resources.
On the other hand 2D and 3D models approach closer real-world phenomena, 
while the observed patterns present more spectacular
and intricate complexity in two (and higher) spatial dimensions
and, often, they do not have equivalent patterns in 1D.

\par As will be explained more extensively
in the next section, the proposed coupled LIF model contains only 
non-local diffusive coupling terms which could account for gap-junction
interactions. Examples of these are the interactions between interneurons
 (neurons that connect other neurons with the central nervous system);
 they indeed interact via gap junctions facilitating the synchronous 
firing of neuronal networks. This synchronization is essential in 
various brain functions, such as oscillatory activities (e.g. gamma oscillations) 
and coordinated motor activities.
Chemical interactions are completely absent in this study which,
therefore, focuses on a restricted special case of neuronal interactions 
\cite{mirollo:1990,kuramoto:1991}.

\par
Our work is organized as follows: 
In the next section, Sec. \ref{sec:model}, we recapitulate the main properties of  
the uncoupled LIF model in Sec.~\ref{sec:model-uncoupled} , while in Sec.~\ref{sec:model-2D} we introduce
the coupled LIF dynamics in 2D, which gives rise to bump states
for positive coupling strengths. In Sec.~\ref{sec:quantitative}
we propose quantitative measures which distinguish the moving bump states (where the 
asynchronous domains travel in the network) from the spatially localized ones.
In Sec.~\ref{sec:refractory} we introduce a refractory period in the LIF nodes and
show that refractoriness fixes the positions of the traveling asynchronous 
bumps.
 In Sec.~\ref{sec:idle} we introduce an alternative mechanism based on the presence
of a number of  randomly distributed permanently
idle nodes in the network and we numerically show that it leads to localization
of the asynchronous domains. 
We explore the parameter regions (density of idle nodes and coupling strength)
and find the parameter ranges where stabilization of the bumps' positions occurs.
Although most of our results are obtained by numerical simulations,
in Sec.~\ref{sec:analytical} we propose a more rigorous analytical approach
to study the properties of stationary patterns
in the continuum limit case of our model.
In the concluding Sec.~\ref{sec:conclusions}, we summarize our main results,
outline their possible applications, and discuss open problems.

\section{The Model}
\label{sec:model}
The Leaky Integrate-and-Fire (LIF) model is an extension of the original Integrate-and-Fire (IF) model,
first introduced by Louis Lapicque in 1907, and is based on experimental observations on the giant
squid axon \cite{lapicque:1907,lapicque:1907b,lapicque:1907a}. 
With respect to IF, the LIF  model includes a leakage term
which prohibits the potential to grow to arbitrarily large values,
is versatile and is widely used in computational neuroscience \cite{gerstner:2002}.
In the rest of this section, we first describe the dynamics of a single LIF neuron
with and without a refractory period, in Sec.~\ref{sec:model-uncoupled}.
Then, we present a 2D LIF network with nonlocal coupling, in Sec.~\ref{sec:model-2D},
and describe the different bump states this network supports,
in Sec.~\ref{sec:quantitative}.

\subsection{The uncoupled LIF model with refractoriness}
\label{sec:model-uncoupled}

\par The general form of the uncoupled LIF model consists of one differential 
equation and two event-driven algebraic conditions. It describes 
the dynamics of the potential $u(t)$ of a single, uncoupled neuron \cite{tsigkri:2017}:

\begin{subequations}
\begin{equation}
 \frac{du(t)}{dt}= \mu-u(t)  
\label{eq01a} 
\end{equation}
\begin{equation}
 \lim_{\epsilon \to 0}u(t+\epsilon ) \to u_0,  \>\>\> {\rm when} \>\> u(t) \ge u_{\rm th} 
\>\> \boldsymbol{\mid} \> {\rm with} \>\> u_{\rm th} < \mu,
\label{eq01b}
\end{equation}
\begin{equation}
 u(t)\equiv u_0,  \> \forall t : [l (T_\mathrm{s}+T_\mathrm{r})+T_\mathrm{s}] \le t \le [(l+1)(T_\mathrm{s}+T_\mathrm{r})],\>\>\>
 {\rm where} \>\> l=0,1,2,\cdots .
\label{eq01c}
\end{equation}
\label{eq01}
\end{subequations}
\noindent 
Equation~\eqref{eq01a} describes the integration of the potential $u(t)$ and the parameter $\mu$ represents the
integration rate. The leakage is here introduced by the term $-u(t)$ which prohibits the potential
$u$ to increase to infinite values, but tends asymptotically to $\mu$ in the absence of condition Eq.~\eqref{eq01b}. 
 Condition~\eqref{eq01b} accounts for the neuron 
firing process and expresses the resetting of the potential to its ground value $u_0$  every time  $u(t)$ reaches the threshold potential $u_{\rm th}$. 
This resetting condition is essential in the LIF model 
because it drives the system to oscillatory behavior. In the absence of 
condition~\eqref{eq01b}
the LIF model is driven to the fixed 
point $u(t)=\mu$ at the asymptotic limit $t\to\infty$.
The threshold potential respects always the inequality 
$u_{\rm th} < \mu$, because $u(t)$ approaches the limiting value $\mu$ only at the asymptotic limit.

\par  In the absence of refractoriness (Eq.~\eqref{eq01c}),
a single neuron described by Eqs.~\eqref{eq01a} and~\eqref{eq01b} behaves as a relaxation oscillator.
Its state before reset as well as the period $T_\mathrm{s}$ can be found as follows:
Assuming that at $t = 0$ the potential is at the rest state, $u_0$, Eq.~\eqref{eq01a} can be solved analytically.
 The solution is $u(t)=\mu -(\mu-u_0)e^{-t}$ for $u(t)\le u_{\rm th}$. Then,
 $T_s$ is obtained via the condition $u(T_\mathrm{s})=u_{\rm th}$, or in other words at $t=T_\mathrm{s}$ the potential
has reached the threshold. Therefore, the period of the single LIF element is given in terms of the
system parameters $(\mu, u_0, u_{\rm th})$ as:
\begin{eqnarray}
 T_\mathrm{s}=\ln \left[ (\mu -u_0)/(\mu - u_{\rm th})\right] .
\label{eq02}
\end{eqnarray}

\par The third condition, Eq.~\eqref{eq01c}, expresses the property of refractoriness 
\cite{gerstner:2002,ermentrout:1998}. After firing, each neuron spends a refractory period $T_\mathrm{r}$ at the
rest state potential $u_0$. During this period the potential does not integrate and in the case of
coupled neurons the connections become non-functioning. Consequently, the period $T$ is the sum of the single
neuron period plus the refractory period, $T=T_\mathrm{s}+T_\mathrm{r}$. In Eq.~\eqref{eq01c}, the index $l$ accounts 
for the successive refractory periods in the evolution of the neuron potential.

\par Using the formula for the period, Eq.~\eqref{eq02}, it is possible to define the 
firing rate $f_\mathrm{s}$ of the single LIF oscillator with and without refractory period, which will be used in later sections. 
Because neurons fire once when the threshold potential $u_{\rm th}$ is reached, $f_\mathrm{s}$ 
 is defined as the inverse of the period:
\begin{eqnarray}
 f_\mathrm{s}=1/(T_\mathrm{s}+T_\mathrm{r}).
\label{eq022}
\end{eqnarray}
\par This definition holds equally for the case of a finite refractory period ($T_\mathrm{r} >0$) or in its
absence ($T_\mathrm{r} =0$).
\par The above discussion together with Eqs.~\eqref{eq01} represent the dynamics of a single, uncoupled
neuron. In the next section, ~\ref{sec:model-2D}, coupling is introduced in the 2D system 
via nonlocal links.

\subsection{The coupled nonlocal dynamics in 2D}
\label{sec:model-2D}

When many LIF elements are coupled in a network, a vector potential
 is introduced to describe the state of the system at time $t$. 
In this work, we assume that the network is a discrete 2D torus,
composed by $N \times N$ nodes with periodic boundary conditions
in both $x$- and $y$-directions.
Each node is defined by a pair of indices $(j,k)$
and the corresponding potential is denoted as $u_{jk}(t)$.
Coupling between neurons is written as a linear superposition of these potentials
and can be roughly related to the gap-junction (or electrical synapse) mechanism.
More specifically, assuming common parameters
$( \mu, u_0, u_\mathrm{th}, T_\mathrm{r} )$ to all network nodes,
the network dynamics is described by the following equations/conditions:
\begin{subequations}
\begin{equation}
 \frac{du_{jk}(t)}{dt}= \mu - u_{jk}(t) +\frac{\sigma}{(2R+1)^2}\sum_{m=j-R,\> n=k-R,}^{j+R,\> k+R}
[u_{mn}(t)-u_{jk}(t)] 
\label{eq03a} 
\end{equation}
\begin{equation}
 \lim_{\epsilon \to 0}u_{jk}(t+\epsilon ) \to u_0,  \>\>\> {\rm when} \>\> u_{jk}(t) \ge u_{\rm th} 
\>\> \boldsymbol{\mid} \> {\rm with} \>\> u_\mathrm{th} < \mu,
\label{eq03b}
\end{equation}
\begin{equation}
u_{jk}(t)\equiv u_0,  \>\>\>  \forall t : T_{jk}^{(p)} \leq t \leq T_{jk}^{(p)} + T_\mathrm{r}
\label{eq03c}
\end{equation}
\begin{equation}
 u_{mn}(t)\equiv u_0,  \>\>\> \forall (m,n) \in \boldsymbol{A}.
\label{eq03d}
\end{equation}
\label{eq03}
\end{subequations}
Here, periodic boundary conditions in the $x$- and $y$-directions mean
that all indices $j,k,m,n$ in Eq.~\eqref{eq03} are considered $\mod N$.
Moreover, since the range of connectivity is $R$ in both directions,
each node is connected with all other nodes in a square of linear size (2R+1) surrounding it.
For normalization purposes the sum on the right-hand size of Eq.~\eqref{eq03a} is divided
by the number of nodes included in this square, equal to $(2R+1)^2$.
In Eq.~\eqref{eq03a}, two network parameters are introduced
in addition to the dynamics parameters,
namely $\sigma$ denoting the coupling strength and $R$ the coupling range.
Both network parameters, $(\sigma , R)$, as well as the node parameters,
$(\mu , u_0, u_\mathrm{th},T_\mathrm{r})$, are common to all elements.
Equation~\eqref{eq03b} represents the resetting condition.
We note here that although the condition is identical for all nodes, each node may fire
at a different time instance due to the influence of the neighbors. Similarly, for the refractory
period implemented by Eq.~\eqref{eq03c}: If we denote as $T_{jk}^{(p)}$ the $p$-th firing of 
the node $(j,k)$, condition Eq.~\eqref{eq03c} states that the node $(j,k)$ remains at the 
rest state $u_0$ for a time period $T_\mathrm{r}$ after firing. Note also that, unlike the single 
uncoupled LIF neuron, the coupled LIF neurons do not have regular (constant) spiking periods
due to the coupling with the neighbors. 
\par In addition to the three equations governing the single neuron, 
a fourth condition is here
imposed concerning the presence of $\boldsymbol{n}$
 randomly distributed idle elements in the system.
Call $\boldsymbol{A}$ the  set of randomly distributed idle elements. The elements
$k=1,2\cdots, \boldsymbol{n}$, 
of set $\boldsymbol{A}$ are randomly chosen at the beginning of the dynamics and do not change during the course of 
the simulation, neither in position nor in potential. 
The nodes of the predefined idle set $\boldsymbol{A}$ are also defined based
on a pair of coordinates $(m,n)$.  
Equation~\eqref{eq03d} refers to these randomly selected elements of set $\boldsymbol{A}$ which always maintain
their rest state, $u_{m n}(t)=u_0$, for all times and for all pairs $(m,n)$ belonging to set $\boldsymbol{A}$.
The percentage of idle or ``dead'' nodes in the system is symbolized by the letter $d$ (for ``dead'') 
and is equal to $d=\boldsymbol{n}/N^2$, where $\boldsymbol{n}$ denotes the number of elements or cardinality
of set $\boldsymbol{A}$.
 In particular simulations, the last two conditions will be relaxed and the cases $\boldsymbol{A} \equiv \emptyset$ or $T_\mathrm{r}=0$ 
will be also discussed.

\subsection{Moving and spatially localized bump states}
\label{sec:quantitative}

In this section, we describe two main classes of bump states  
found in the system~\eqref{eq03}, namely, the moving bumps and 
the spatially localized ones.
Several quantitative measures are also proposed to characterize them.
For the production of these patterns
numerical integration of Eq.~\eqref{eq03} was performed
using the forward Euler method with a fixed time step 
$\Delta t=0.001$.
All simulations started from random initial conditions,
where each $u_{jk}$ was chosen randomly and independently
from the interval $[u_0,u_\mathrm{th})$.
In most of the cases discussed below,
we used the following set of default parameters:
$\mu=1.0$, $u_\mathrm{th}=0.98$, $u_0=0$, $\sigma =0.7$ and $N\times N=64 \times 64$.
In addition, equivalent systems of size $32 \times 32$ were used
to speed up the calculation of two-parameter scans
in Sec.~\ref{sec:refractory} and Sec.~\ref{sec:idle},
while in Sec.~\ref{sec:analytical} systems of size up to $256 \times 256$
were employed to address the continuum limit behavior.

\par
\begin{figure}[h]
\includegraphics[height=0.21\textwidth]{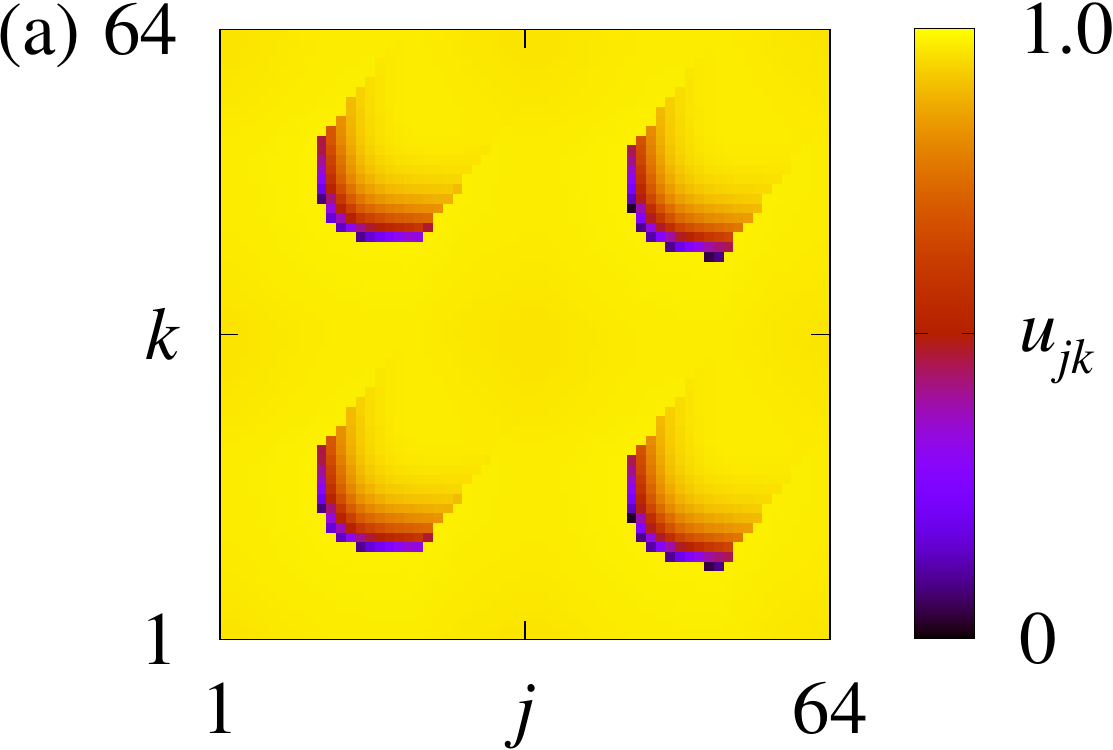}\hspace{2mm}
\includegraphics[height=0.21\textwidth]{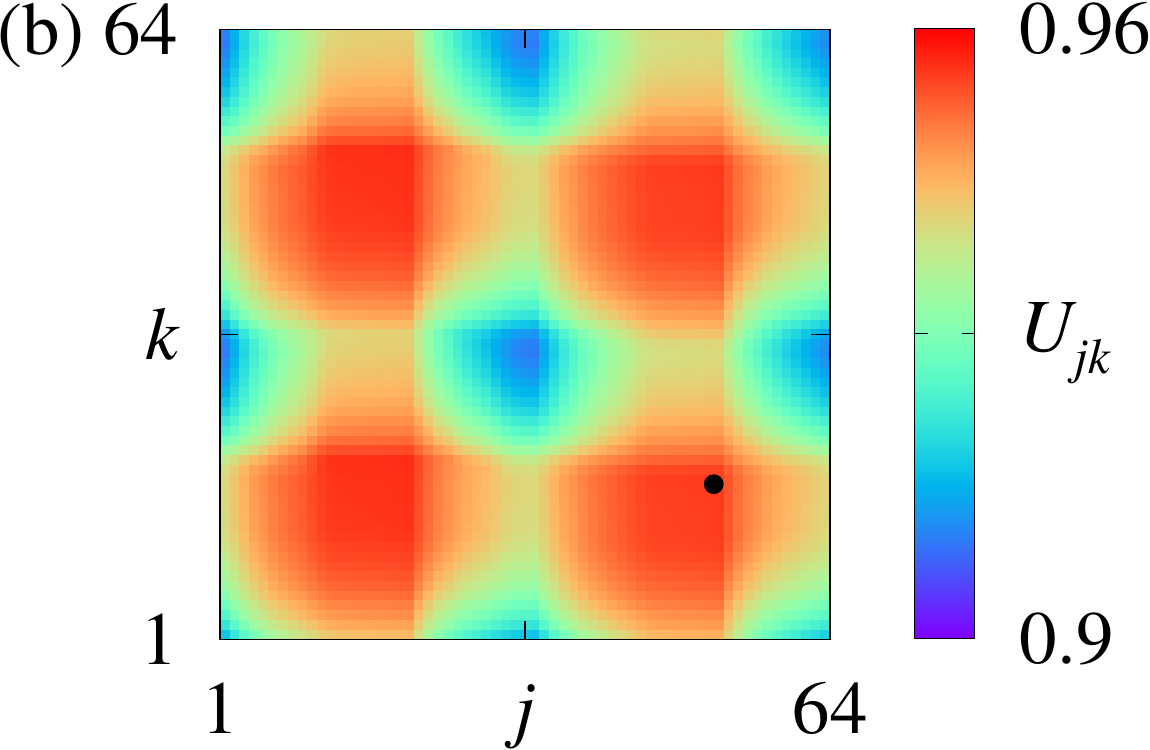}\hspace{2mm}
\includegraphics[height=0.21\textwidth]{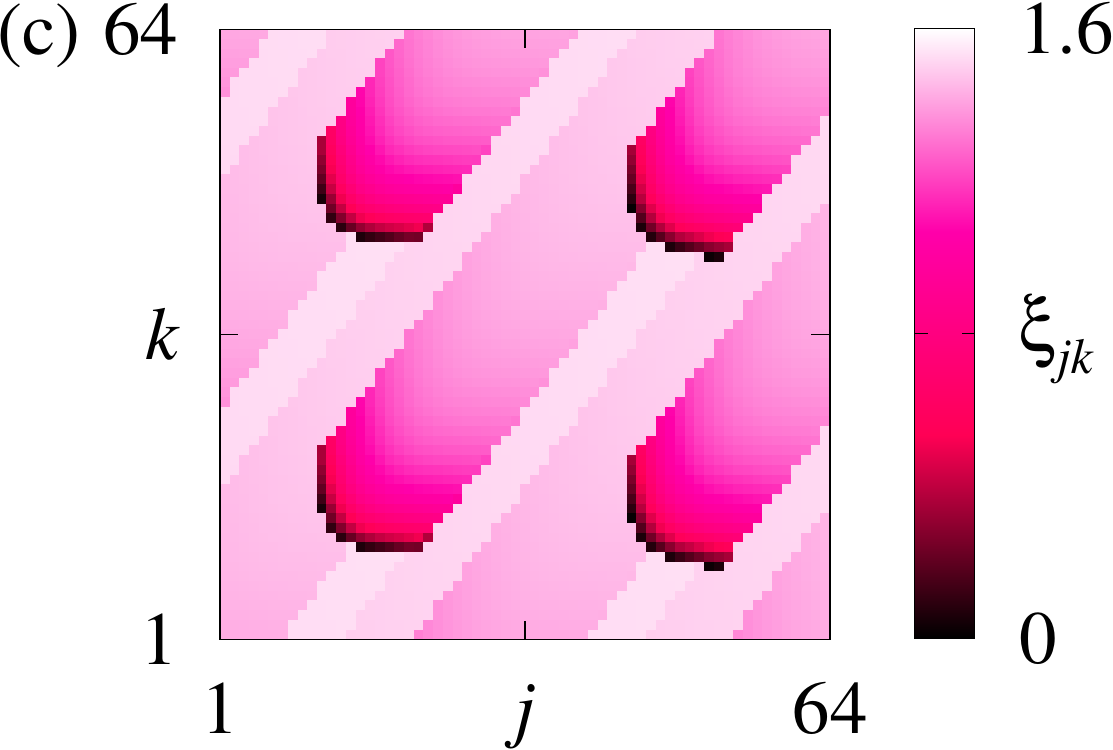}\\[3mm]
\includegraphics[height=0.20\textwidth]{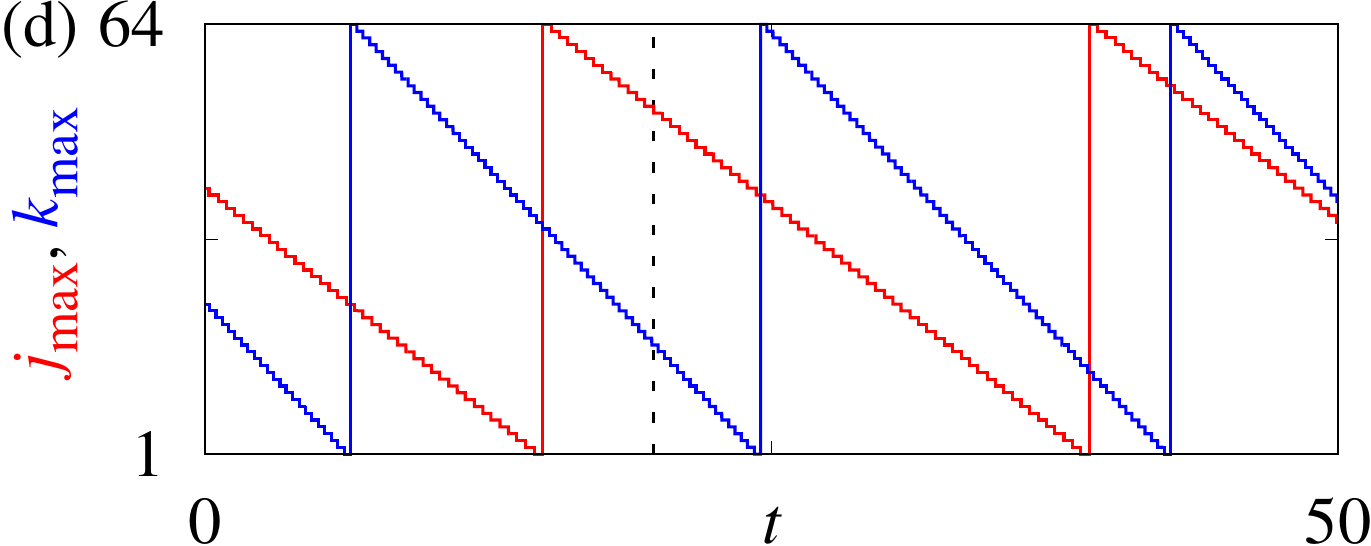}
\caption{\label{fig:Tr0_0_}
(Color online) 
Moving bump state in the LIF model~\eqref{eq03}:  
(a) the potential $u_{jk}$, (b) the mean field potential $U_{jk}$,
and (c) the firing history function $\xi_{jk}$ of the neurons.
(See related video: fig1\_movie.gif).
The bullet in panel (b) indicates the bump position
determined as the maximum of the $U_{jk}$-profile.
Panel (d) shows the bump position as a function of time.
The panels (a-c) correspond to the time moment
indicated by the black dashed line in panel (d).
Parameters: $\mu=1.0$, $u_{\rm th}=0.98$, $u_0=0$, $\sigma=0.7$,
$R=22$, $N\times N=64 \times 64$, $T_\mathrm{r} = 0$ and $d=0$.
All simulations start from random initial conditions.
}
\end{figure}

\begin{figure}[h]
\begin{center}
\includegraphics[height=0.21\textwidth]{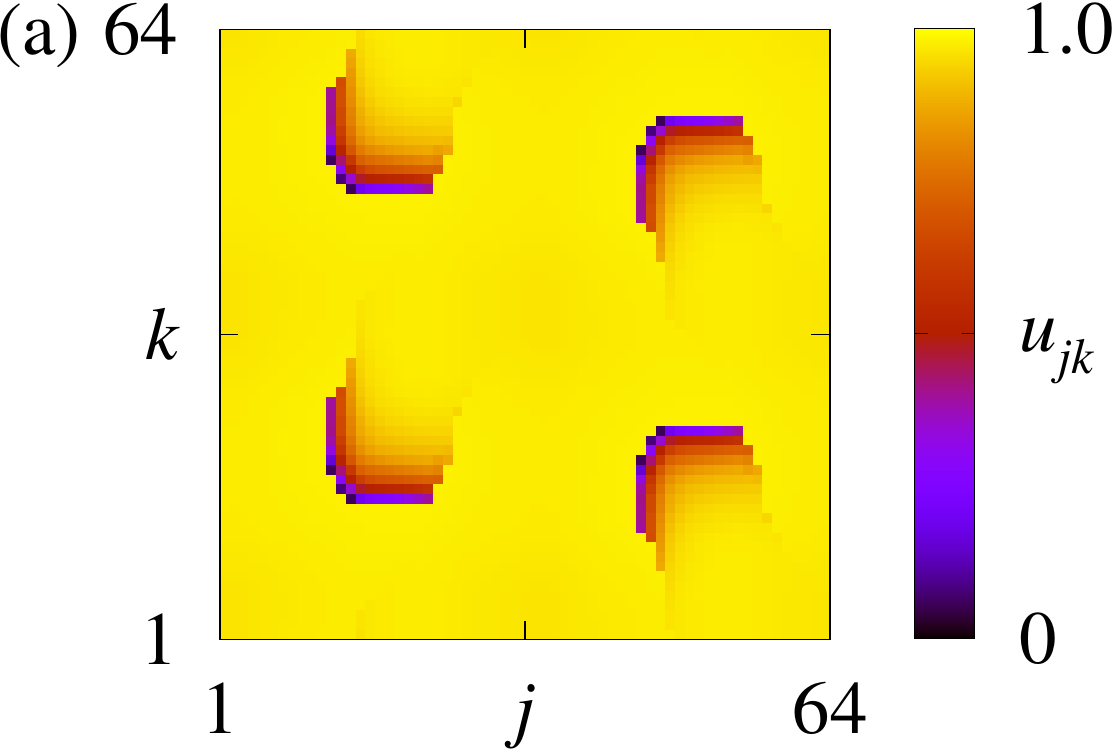}\hspace{2mm}
\includegraphics[height=0.21\textwidth]{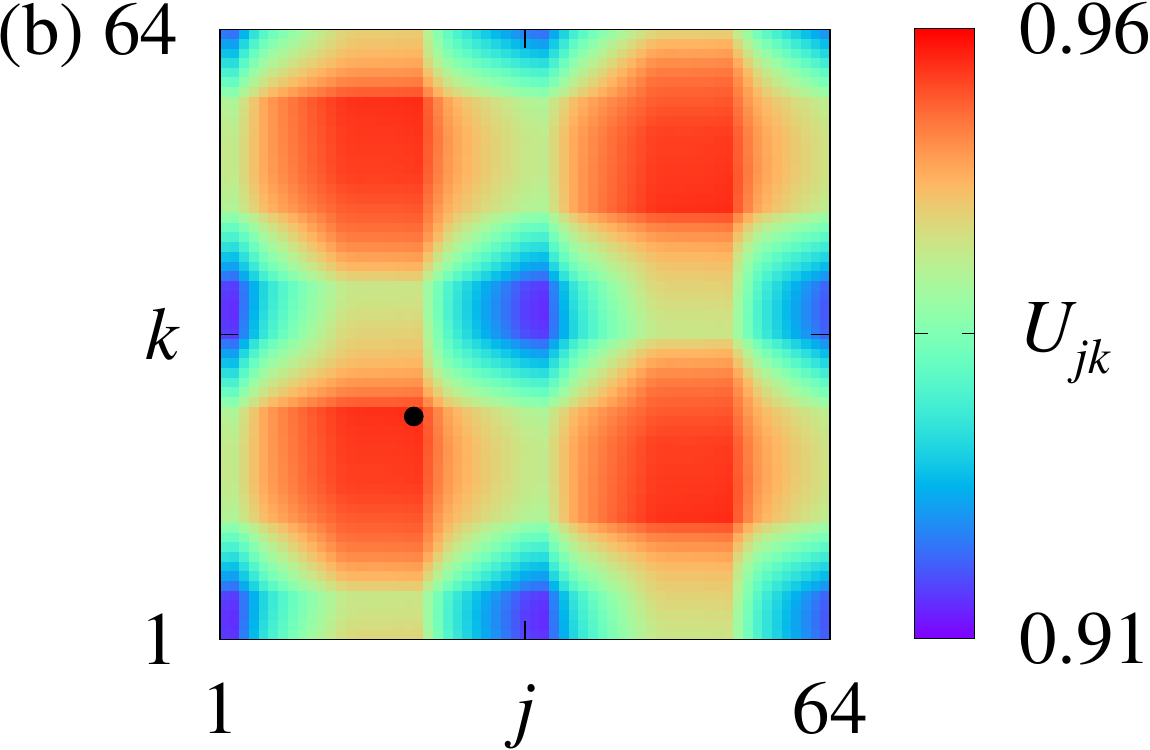}\hspace{2mm}
\includegraphics[height=0.21\textwidth]{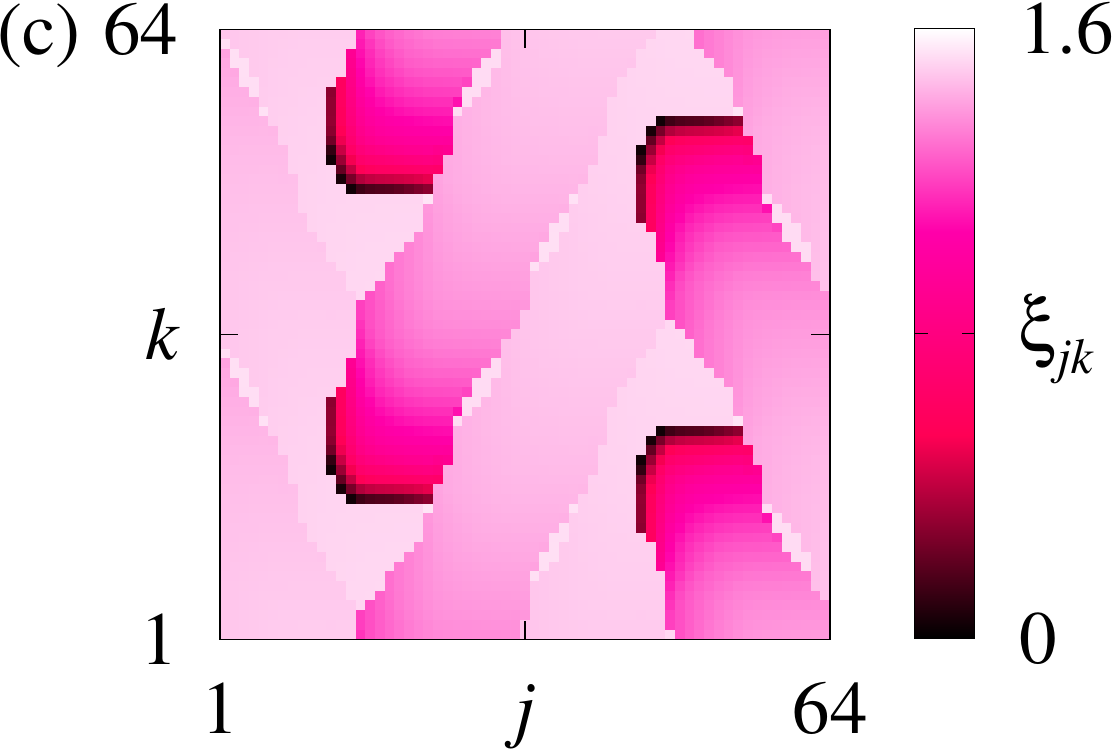}\\[2mm]
\includegraphics[height=0.21\textwidth]{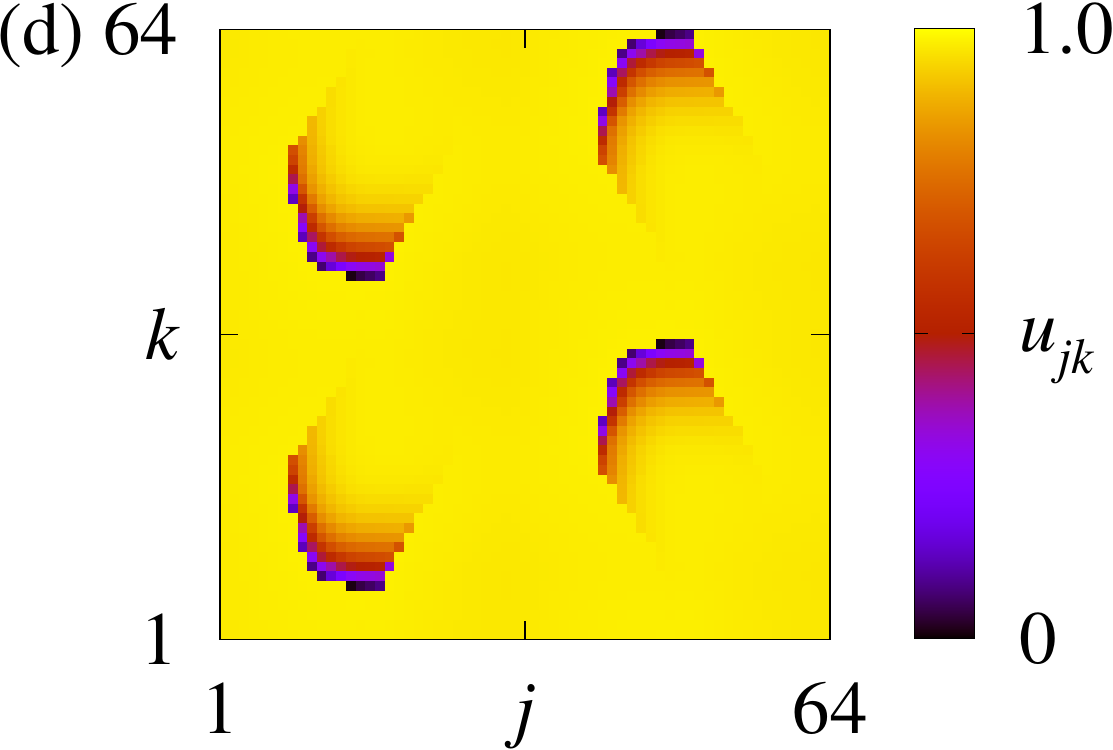}\hspace{2mm}
\includegraphics[height=0.21\textwidth]{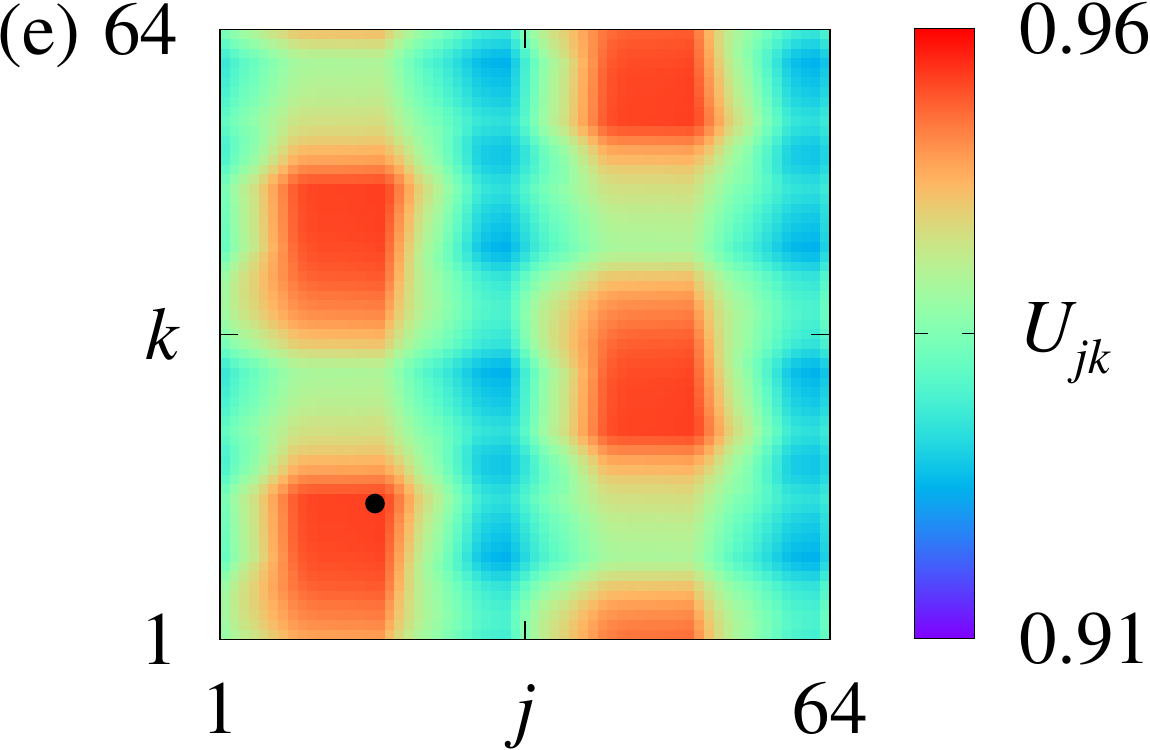}\hspace{2mm}
\includegraphics[height=0.21\textwidth]{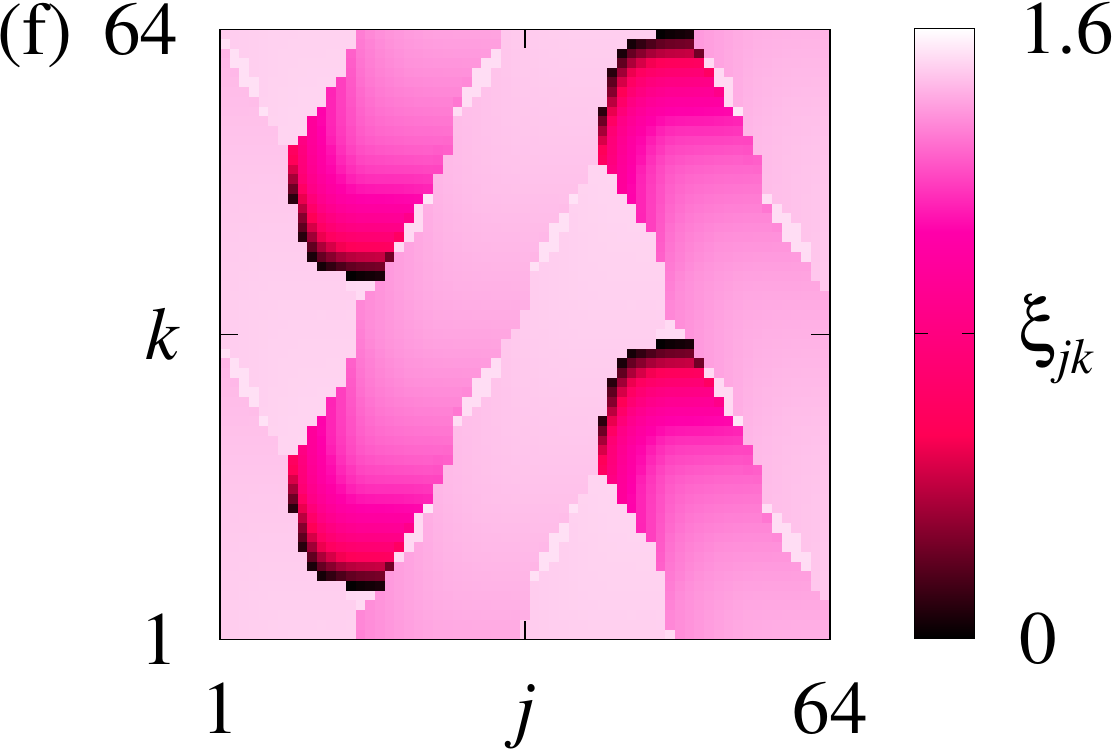}\\[3mm]
\includegraphics[height=0.20\textwidth]{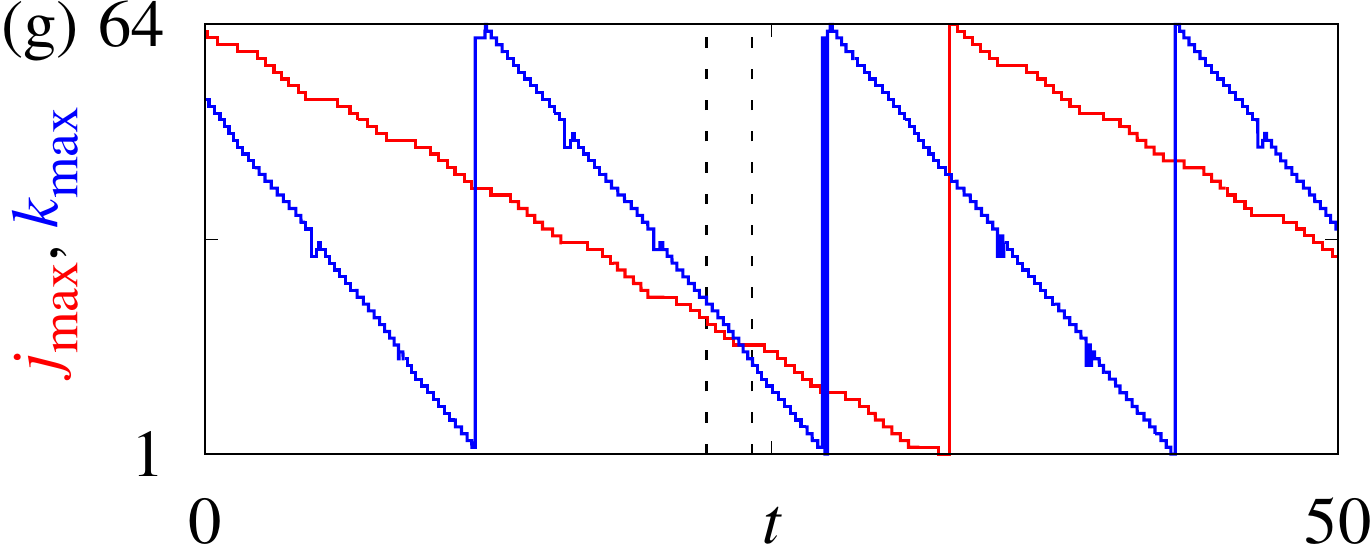}
\end{center}
\caption{\label{fig:Tr0_0}
(Color online) Second type of moving bump state coexisting with the one shown in Fig.~\ref{fig:Tr0_0_}. (See related video: fig2\_movie.gif). 
The panels (a-c) and (d-f) correspond to the two consecutive times
indicated by the dashed lines in panel (g).
}
\end{figure}

 In Figs.~\ref{fig:Tr0_0_} and~\ref{fig:Tr0_0}, we show two moving bump states
that stably coexist in Eq.~\eqref{eq03} for $R = 22$ and $N\times N=64 \times 64$,
in the absence of idle neurons ($d = 0$) and with
 vanishing refractory period ($T_\mathrm{r} = 0$).
These states were obtained using different realizations
of random initial conditions for $u_{jk}(t=0)$,
keeping all other system parameters identical.
In addition to the snapshot $u_{jk}$ for each pattern
we also show its {\it mean field potential}
\begin{equation}
U_{jk}(t) = \frac{1}{(2R+1)^2} \sum_{m=j-R,\> n=k-R}^{j+R,\> k+R} u_{mn}(t).
\label{Def:U}
\end{equation}
The mean filed potential has a pronounced four-hump profile
as evidenced in Figs.~\ref{fig:Tr0_0_}(b) and ~\ref{fig:Tr0_0}(b),
so that the position of one of these humps
(more specifically the position of its local maximum
$( j_\mathrm{max}, k_\mathrm{max} )$)
can be used as a marker to track the movement of the pattern.

\par Another way of visualizing pattern dynamics
relies on the firing history of different neurons.
Indeed, for each neuron at time $t$, we can determine the time interval
that has passed since its last firing $\tau_{jk}(t)$.
(If the neuron has not yet fired at all, we assume $\tau_{jk}(t) = t - t_0$
where $t_0$ is the initial moment of simulation.)
From the $\tau_{jk}(t)$-snapshot, we can infer the firing order of different neurons
and thus recognize the appearance of traveling waves or more complex moving patterns.
However, the value $\tau_{jk}(t)$ is not entirely convenient, because it is unbounded.
In particular, for the permanently idle neurons, it grows to infinity over time.
To avoid this complication, we introduce a smooth cut-off
by defining the {\it firing history function} $\xi_{jk}(t) = \arctan \tau_{jk}(t)$.
Examples of such firing history functions
are shown in Fig.~\ref{fig:Tr0_0_}(c) and Fig.~\ref{fig:Tr0_0}(c,f).
Remarkably, they reveal a qualitative difference between these two moving patterns.
Although in both cases we see moving bump states in the form of curved excitation fronts,
the former one propagates at an almost constant speed along a straight line,
while the latter  changes its shape and wiggles as it moves,
compare Fig.~\ref{fig:Tr0_0}(a-c) and Fig.~\ref{fig:Tr0_0}(d-f). 
Moreover, the latter pattern is characterized by two spatial periods, 
as it consists of two ``groups'' of bumps moving in perpendicular directions, 
however never colliding.

\begin{figure}[h]
\begin{center}
\includegraphics[height=0.21\textwidth]{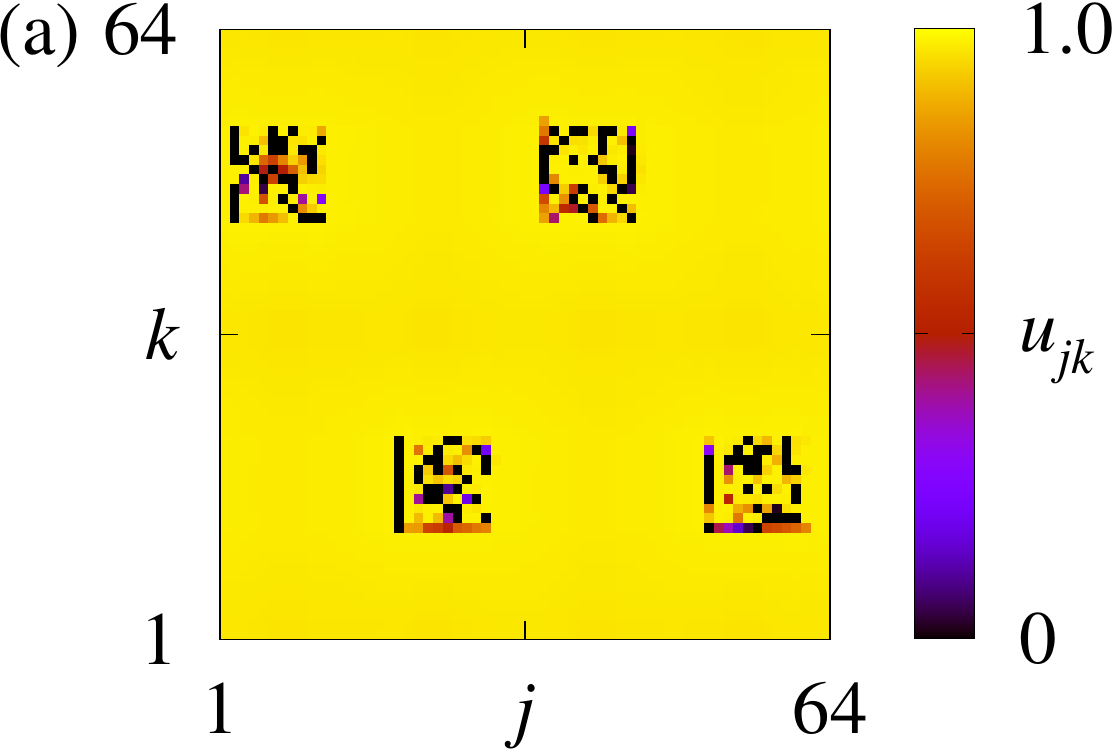}\hspace{2mm}
\includegraphics[height=0.21\textwidth]{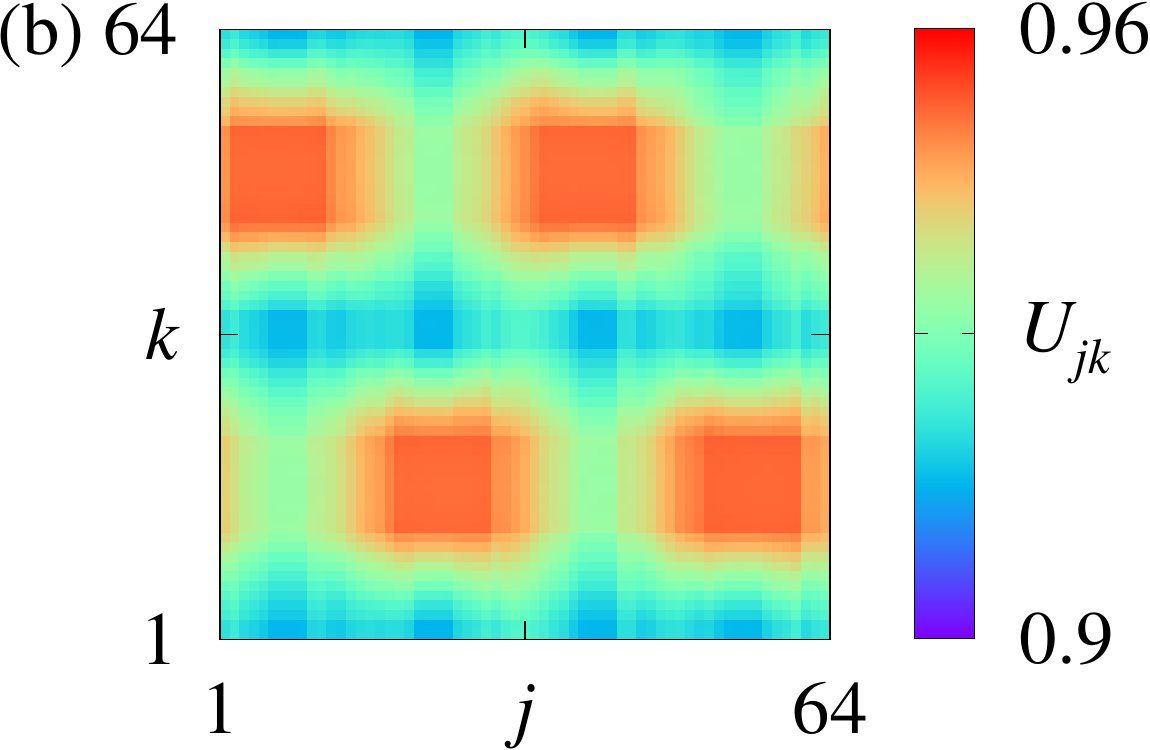}\\[2mm]
\includegraphics[height=0.21\textwidth]{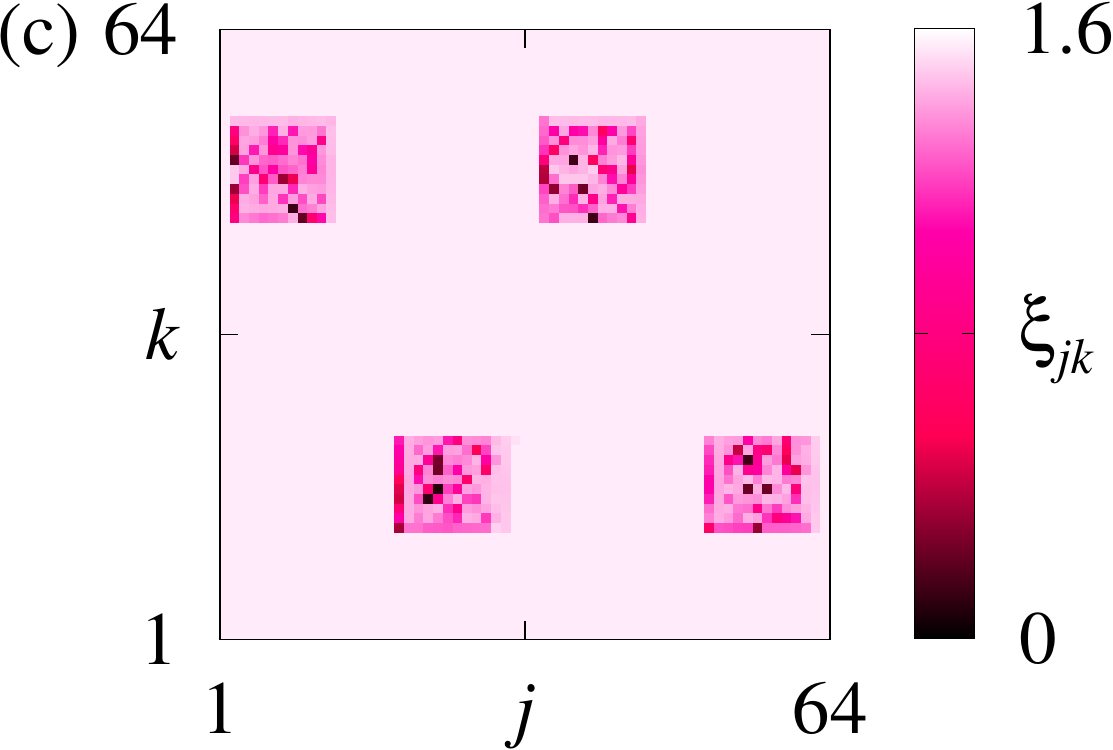}\hspace{2mm}
\includegraphics[height=0.21\textwidth]{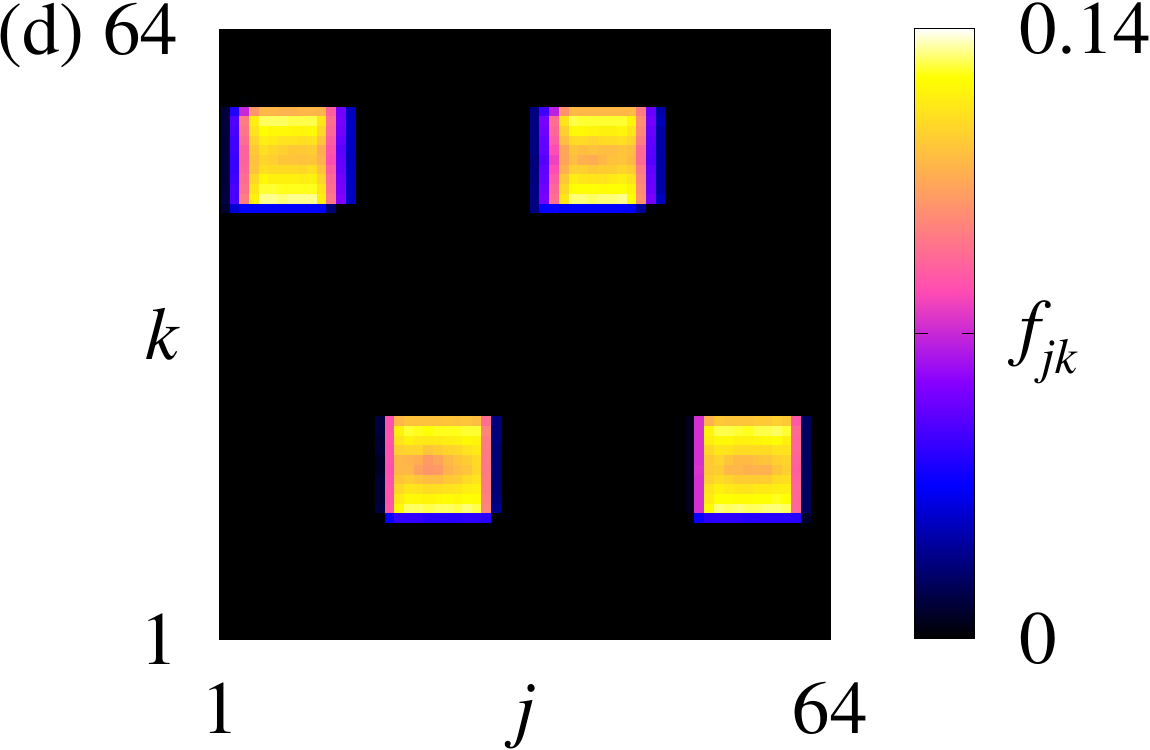}
\end{center}
\caption{\label{fig:Tr2_5}
(Color online)
Spatially localized pattern in the LIF model~\eqref{eq03}:
(a) the potential $u_{jk}$, (b) the mean field potential $U_{jk}$,
(c) the firing history function $\xi_{jk}$,
and (d) the firing rates $f_{jk}$ of the neurons. (See related video: fig3\_movie.gif).
All parameters are the same as in Fig.~\ref{fig:Tr0_0_}, except $T_\mathrm{r} = 2.5$.
The time interval for calculation of $f_{jk}$ is $\Delta T = 1000$.
}
\end{figure}

\par To make the contradistinction to moving bump
states, in Fig.~\ref{fig:Tr2_5}, we show an example of a spatially localized pattern
(the specific parameter values under which this pattern is obtained will be discussed
later, in Sec.~\ref{sec:refractory}).
Its snapshot appears as four structured active regions
embedded in a homogeneous quiescent background. 
When comparing Fig.~\ref{fig:Tr2_5}(c) with Fig.~\ref{fig:Tr0_0_}(c)
and Fig.~\ref{fig:Tr0_0}(c,f), we recognize that the firing of neurons
occurs exclusively in these active regions.
This fact becomes even clearer, if we plot the {\it firing rate} of each neuron
defined by the formula
\begin{eqnarray}
f_{jk}=Q_{jk}/\Delta T,
\label{eq04}
\end{eqnarray}
where $Q_{jk}$ is the number of resettings/firings
of neuron at position $(j,k)$  during the time interval $\Delta T$.
Fig.~\ref{fig:Tr2_5}(d) demonstrates that $f_{jk}$ remains finite
only around the positions of the four specific active regions,
while $f_{jk} = 0$ on the rest of the plane.

For localized bump states, the spatial structure of the firing rate profile $f_{jk}$
can be further analyzed using the following global characteristics~\cite{omelchenko:2013}.
If we calculate the minimal and maximal firing rates:
$$
f_{\rm min}=\min\limits_{j,k} f_{jk}
\quad\mbox{and}\quad
f_{\rm max}=\max\limits_{j,k} f_{jk},
$$
then the difference $\Delta f = f_{\rm max}-f_{\rm min}$
gives a rough estimate of the spatial heterogeneity of the pattern.
Based of the values of $f_{jk}$
it is  possible to define the activity ${\alpha}$ in the network,
provided that the bumps stay mostly immobile.
Let us call $N_\mathrm{a}$ the number of active elements,
i.e., the number of nodes with non-zero $f_{jk}$.
Then the activity in the network may be expressed
as the percentage of active nodes over the total number of nodes,
$\alpha =N_\mathrm{a}/N^2$.

In the rest of the manuscript, we  analyze the effect of the model parameters,
such as the coupling range $R$, the refractory period $T_\mathrm{r}$
(in Sec. \ref{sec:refractory}),
and the fraction of permanently idle nodes $d$ (in Sec. \ref{sec:idle}),
on the type of predominant patterns observed in our system.

\section{Mechanism I: Refractory Period}
\label{sec:refractory}

After firing, biological neurons are known to spend a period of time
during which they are hardly excitable,
even if the sum of input signals from connected neurons exceeds the threshold.
This period is called a refractory period, $T_\mathrm{r}$,
and its actual value depends significantly on the type of nerve cell~\cite{gerstner:2002}.
In this section, we take into account the refractory period
and consider the case where, after firing, the neurons spend a time $T_\mathrm{r}$
at the rest state $u_0$ (see Eq.~\eqref{eq03c}).
In addition, all neurons are assumed to be in the active mode
and none of them is preset in a damaged or permanently idle state.
Equivalently, $\boldsymbol{A}\equiv\emptyset$
and $\boldsymbol{n}(\boldsymbol{A})=0$ in Eq.~\eqref{eq03d}.

\par A rough idea of the influence of the refractory time $T_\mathrm{r}$
on the bump states in system~\eqref{eq03} can be obtained from comparison of
 Fig.~\ref{fig:Tr0_0_} and Fig.~\ref{fig:Tr2_5}.
Recall that in the first case the refractory period
$T_\mathrm{r} = 0$ is set, and in the second case $T_\mathrm{r} = 2.5$,
with all other parameters and initial conditions being identical in these figures.
For comparative reasons we provide here the firing rates for the uncoupled
neurons in the case of no-refractory period as in Figs.~\ref{fig:Tr0_0_},~\ref{fig:Tr0_0} and for $T_\mathrm{r} = 2.5$
as in Fig.~\ref{fig:Tr2_5}. 
For the chosen parameters $(\mu, u_0, u_\mathrm{th})$
of individual neurons, formula~\eqref{eq02} gives the period value $T_\mathrm{s} =3.9$,
and the corresponding firing rate (with $T_\mathrm{r} = 0$)
equals $f_\mathrm{s} = 1/T_\mathrm{s} = 0.2564$. With the same parameter values and
$T_\mathrm{r} = 2.5$,   $f_\mathrm{s} = 1/(T_\mathrm{s}+T_\mathrm{r}) = 0.15625$.
\par With respect to firing history plots, comparing the $\xi$-profiles shown in Figs.~\ref{fig:Tr0_0_}(c) and~\ref{fig:Tr2_5}(c),
we conclude that in the 2D-torus network the position of the active spatial regions
gets fixed if a refractory period is added in the dynamics of neural oscillators.
It is then evident that the introduction of a refractory period alone is sufficient to spatially confine the moving patterns.

\subsection{Distinction between moving and spatially localized bump states}

Below, we are interested in distinguishing moving and spatially localized patterns.
For this, we can use, for example, the minimal firing rate $f_{\rm min}$.
Indeed, such a criterion works reliably in the case of 1D ring networks~\cite{tsigkri:2017},
since for localized patterns we should have $f_{\rm min} = 0$,
while for traveling patterns that cross all network nodes
for a sufficiently long time, we get $f_{\rm min} > 0$.
However, this criterion may be false in the 2D case.
If the size of the active region is small enough
and it moves strictly horizontally or vertically
on the flat representation of the torus, see Figs.~\ref{fig:Tr0_0_}-\ref{fig:Tr2_5},
then certain parts of the network remain quiescent and hence $f_{\rm min} = 0$.
Therefore, in the 2D case the presence of elements with $f_{jk}=0$
is a necessary but not sufficient condition to signify pattern localization. 

\par Because of this, we propose another differentiation criterion
involving the maximal firing rate $f_{\rm max}$.
It is based on the following observation.
If a certain pattern does not move in space,
then most neurons in its active region almost maintain their firing rates
and therefore, $f_{\rm max}$ is close to the firing rate of a single neuron,
$f_\mathrm{s}=1 /(T_\mathrm{s}+T_\mathrm{r})$, see Eq.~\eqref{eq022}.
(Usually $f_{\rm max}$ is slightly smaller than $f_\mathrm{s}$,
but in some cases, it can also happen that $f_{\rm max} > f_\mathrm{s}$
due to the network cooperativity effects.)
In contrast, for moving patterns,
the maximal firing rate $f_{\rm max}$ is much smaller than $f_\mathrm{s}$,
because all nodes spend part of their time in the active mode
and part of their time in quiescent mode.
Therefore, it is reasonable to use a comparison of $f_{\rm max}$ with $f_\mathrm{s}$
to distinguish between moving and spatially localized patterns.

\begin{figure}[h]
\includegraphics[clip,width=0.5\linewidth, angle=0]{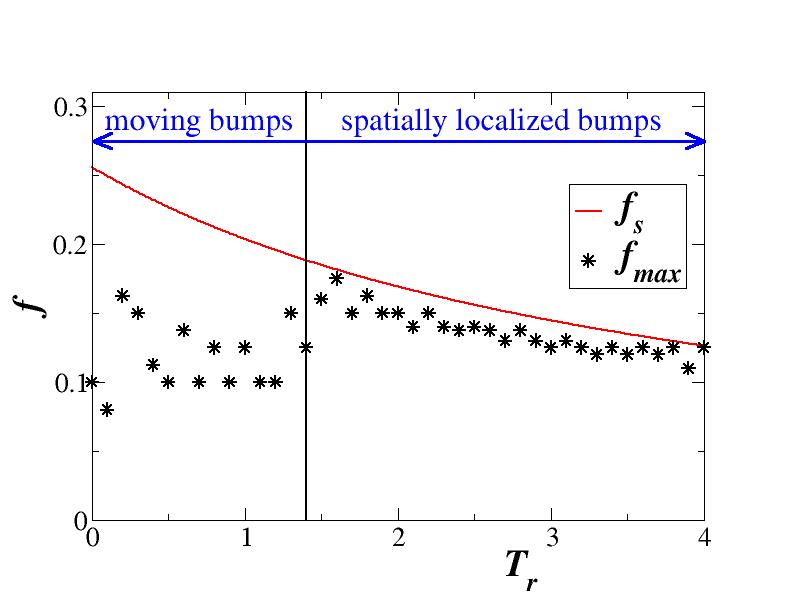}
\caption{\label{fig04} (Color online)
Bump states in the LIF model in 2D. Comparison of $f_{\rm max}$ (stars, from simulations)
 with $f_\mathrm{s}$ (solid red line, from Eq.~\eqref{eq022}).
In the $T_\mathrm{r}$ regions where $f_{\rm max} << f_\mathrm{s}$, i.e., for $T_\mathrm{r} \le 1.4$, the bumps are moving,
otherwise they are confined. The vertical black line at $T_\mathrm{r} = 1.4$ delineates 
the regions of moving and localized bump states.
Parameters are: $\mu=1.0$, $u_{\rm th}=0.98$, $u_0=0$, $\sigma=0.7$, $R=11$, $N\times N=32 \times 32$ and $d=0$. 
}
\end{figure}

For example, in Fig.~\ref{fig04}, we plot the values of $f_\mathrm{s}$
(using Eq.~\eqref{eq022}) and $f_{\rm max}$ as calculated from the simulations,
for comparison. We use $N\times N = 32\times 32$
and a constant value of $R=11$,
where the localized patterns are more abundant.
Other parameters are as in Fig.~\ref{fig:Tr0_0_}.
From a visual inspection of the $\xi$-profiles and the corresponding dynamics movies,
we see that for large values of $T_\mathrm{r}$
the bump states are predominantly localized,
while for smaller values of $T_\mathrm{r}$ they look like traveling excitation fronts.
The transition between these two dynamical regimes occurs at $T_\mathrm{r} = 1.4$,
which is indicated by the vertical black line in Fig.~\ref{fig04}.
We notice that in the left part of Fig.~\ref{fig04},
where the traveling patterns were recorded,
the values of $f_{\rm max}$ fall much lower than $f_\mathrm{s}$,
while for localized patterns on the right part of the plot, we have $f_{\rm max}\approx f_\mathrm{s}$.
(In fact, the recorded $f_{\rm max}$ values stay somewhat below $f_\mathrm{s}$
and this is due to the slight random motion
of the active regions around a central fixed position).
Therefore, plotting the graphs of $f_{\rm max}$ and $f_\mathrm{s}$ together
allows us to effectively distinguish between traveling and spatially localized activity patterns.
In particular, we will use this simple criterion
in Secs.~\ref{Sec:RefractoryPeriodDiagram} and~\ref{sec:idle}.


\subsection{The influence of the refractory period on the moving bump states}
\label{Sec:RefractoryPeriodDiagram}

\begin{figure}[h]
\includegraphics[clip,width=0.48\linewidth, angle=0]{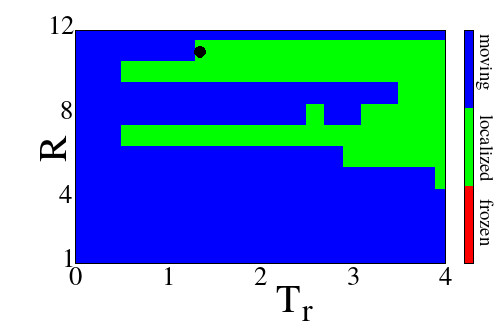}
\caption{\label{fig03} (Color online)
Color-coded map of moving or spatially localized bump states in LIF model with refractory period in 2D. 
Moving bumps are represented by blue color
and spatially localized bumps by green color. 
The black dot denotes the transition from mobile to immobile
bumps at $R=11$ and $T_r=1.4$, for comparison with Fig.~\ref{fig04}.
Frozen states are not observed here but the scale includes the red
color (attributed to potential frozen states) for comparison with Fig.~\ref{fig07}, later on.
Parameters are: $\mu=1.0$, $u_{\rm th}=0.98$, $u_0=0$, $\sigma=0.7$, $N\times N=32 \times 32$ and $d=0$.
All simulations start from the same random initial conditions.
}
\end{figure}

In Fig.~\ref{fig03} we explore the presence of localized and moving bumps
for different values of the coupling range $R$ and refractory period $T_\mathrm{r}$.
In the color-coded plot the moving bumps are represented by the blue color
and the localized or confined ones by the green color.
The mobility of the bumps was here estimated via the firing rate profiles,
as the ones in Fig.~\ref{fig:Tr2_5}(d). Roughly speaking, Fig.~\ref{fig03} shows
that the bumps get spatially confined on the right hand-side of the plot,
where the values of $T_\mathrm{r}$ are large.

\par Note that the bumps we call localized are not necessarily immobile.
In some cases, their active regions occupy a fixed portion of space,
while in other cases these regions move erratically around some central position.
In the color-coded plot, both situations are collectively represented by green color.
Remarkably, the random motion of the bumps can often be attributed to finite size 
effects \cite{compte:2000,cchow:2006},
so that it disappears for large system sizes $N$.
For illustration of the random motion of the bumps 
(under specific parameter values and initial conditions) 
a related video ``fig6a\_movie.gif''
is in the Supplementary Material.
In addition, the case of stationary bumps in the limit $N\to\infty$
will be further considered in Sec.~\ref{sec:analytical}.

\par As a general note on the role of refractoriness, 
we demonstrated numerically that the introduction of a refractory period in 
the dynamics of LIF networks causes localization of the bumps in space, mostly for large refractory periods. 
We also recall that refractoriness is an intrinsic property of biological neuron cells.
Given the above, we may propose that one of the biological reasons of neural 
refractoriness may be the control of the motion of the
traveling bumps carrying and distributing information in the form of potential packets within 
the networks of living neurons. It is important to also emphasize 
that a specific requirement for the presence of traveling or localized bumps is the positiveness of the coupling, which is experimentally observed in the synapses of biological neuron networks.
Although in the network of the healthy brain both positive and negative synapses are
present \cite{brunel:2000,kim:2020}, there are situations where positive synapses (in the form of excitatory interactions)
are important because they facilitate the signal transmission between neurons \cite{hansel:1995}. 
In addition, they play a key role in synaptic plasticity, i. e. the ability of 
synapses to strengthen or weaken over time (not considered in the present study). 
At the same time, they balance inhibitory synapses 
to maintain a proper level of excitability in the brain. On the downside, 
when excitatory synapses dominate due to increased excitatory neurotransmitter 
release (e.g., glutamate) this can result in epileptic seizures \cite{abegg:2004,bod:2023}.
\par 
Overall, our numerical results using the coupled LIF network with positive
couplings indicate, on the one hand,
that bump states are realizable in 
networks with positive coupling strengths and, on the other hand, 
that refractoriness controls the motion/localization of the bumps.

\section{Mechanism II: Permanently Idle Randomly Distributed Nodes}
\label{sec:idle}

\par In this section we introduce an alternative mechanism
which leads to stabilization of the positions of the asynchronous
domains. Namely, we consider the case where the set $\boldsymbol{A}$ is
nonempty, i.e., a number of randomly selected nodes are preset at the
rest state, $u_0$, and they remain idle during the course of the simulation.
This scenario mimics the presence of ``dead'' neuron cells in random locations.
The case where the link weights gradually decay approaching zero
is an altogether different scenario and is relevant in
 neurodegenerative diseases, 
for instance Alzheimer’s disease which is linked to pathological desynchronization 
and decoupling of neuronal populations \cite{mondragon:2019,kromer:2020}. 
The case of link weights decay is beyond the scope of the 
present study.

\par We would like to stress that the preselected nodes were
permanently idle during the entire process, as if
they were blocked, or broken or completely damaged (``dead'').
If the idle nodes are chosen dynamically,
i.e., if different nodes are chosen in each iteration step to remain silent,
our simulations indicate that the asynchronous regions still travel in the network
and do not stabilize in space. This is because the moving bumps do not find permanent
obstacles during their motion but may be blocked only temporarily by the instantaneous 
resetting of the nodes which become ``alive'' directly after resetting and, the bumps
 continue their motion at the next instances.

\par The mobility of the active asynchronous domains in the presence of randomly distributed
idle nodes is here investigated for the 2D LIF network. 
To keep the system as simple as possible and to differentiate
between mechanisms I and II, in the following
we remove the effects of the refractory period and include a small number of permanently idle nodes.
Equivalently, we set $T_\mathrm{r} = 0$ and $d>0$ in Eqs.~\eqref{eq03}. 
\begin{figure}[b]
\begin{center}
\includegraphics[height=0.21\textwidth]{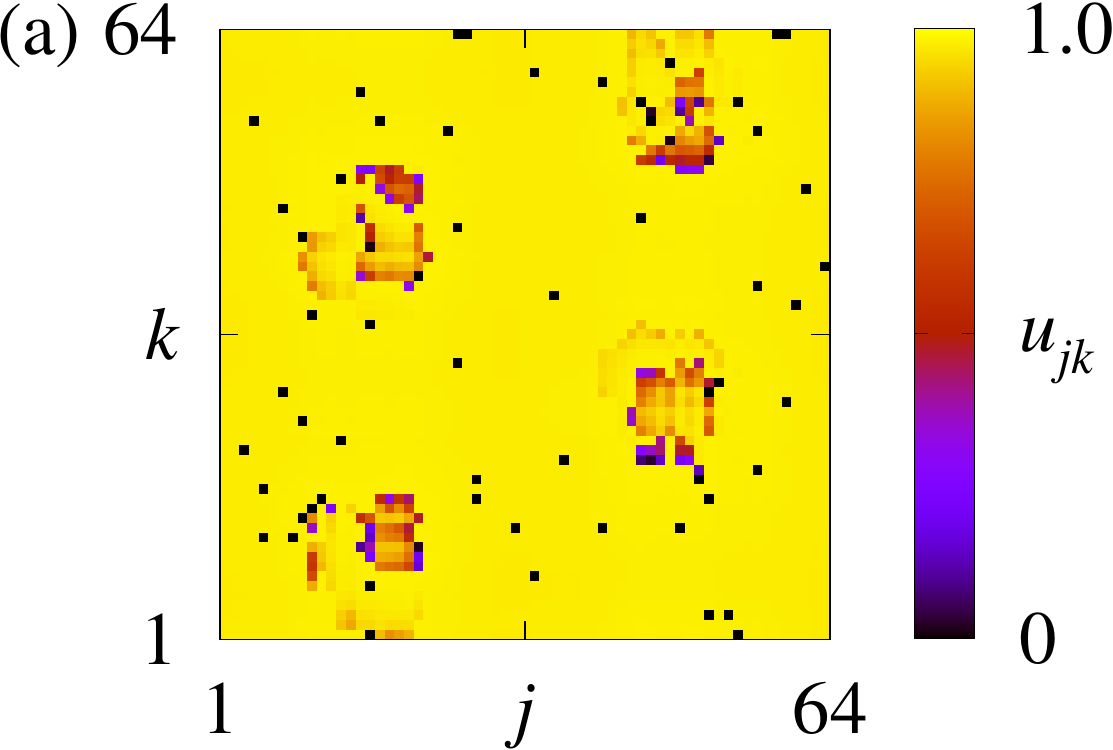}\hspace{2mm}
\includegraphics[height=0.21\textwidth]{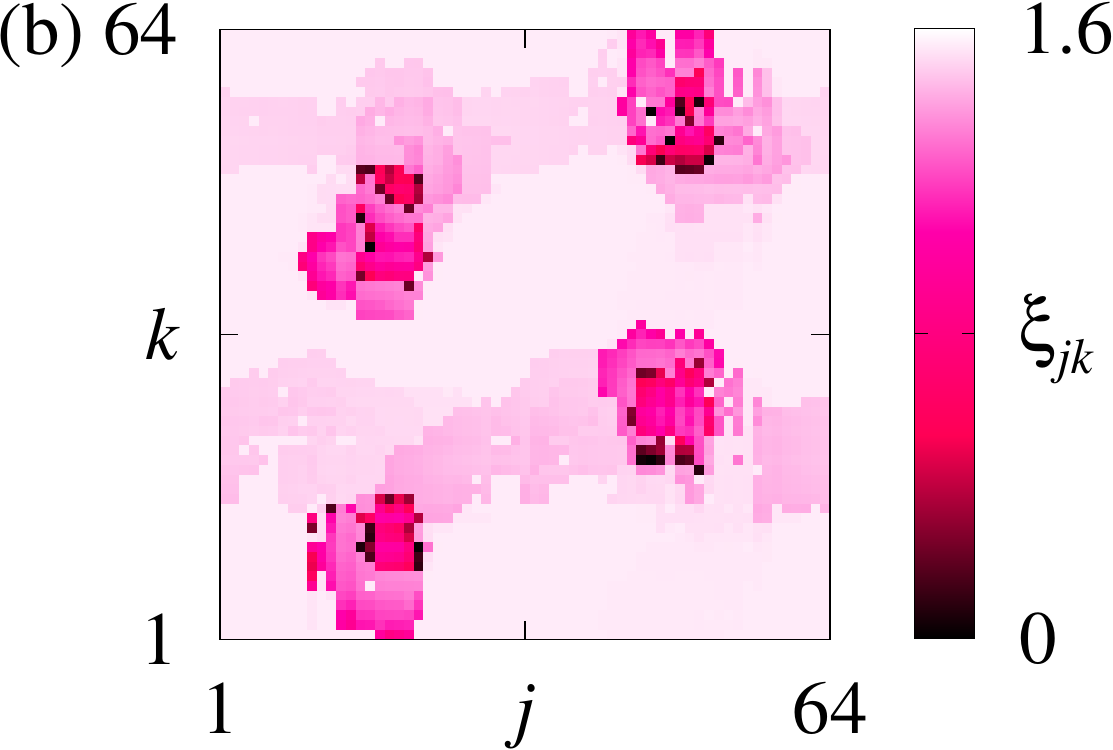}\hspace{2mm}
\includegraphics[height=0.21\textwidth]{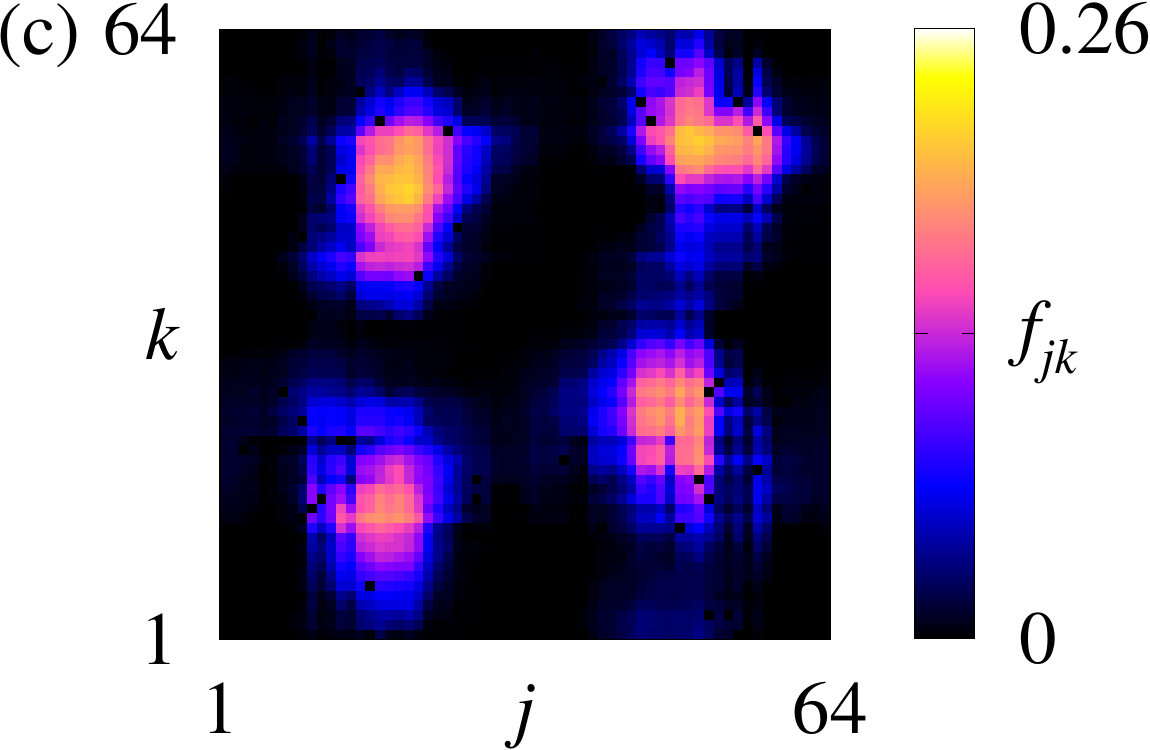}\\[2mm]
\includegraphics[height=0.21\textwidth]{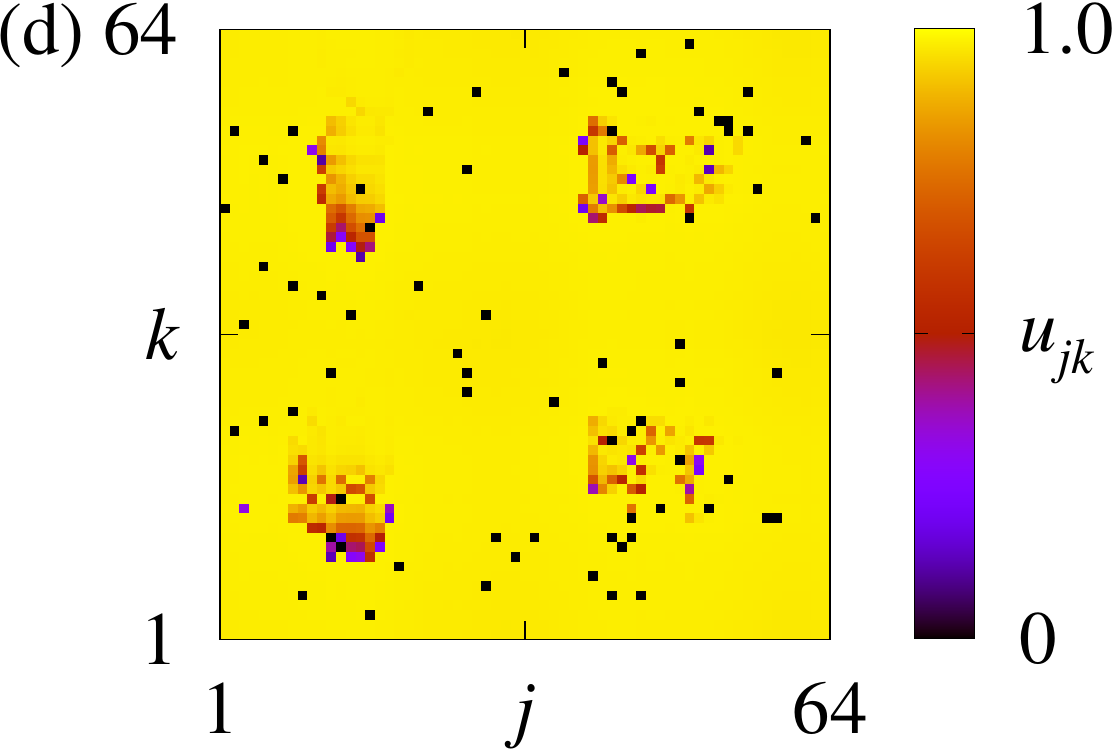}\hspace{2mm}
\includegraphics[height=0.21\textwidth]{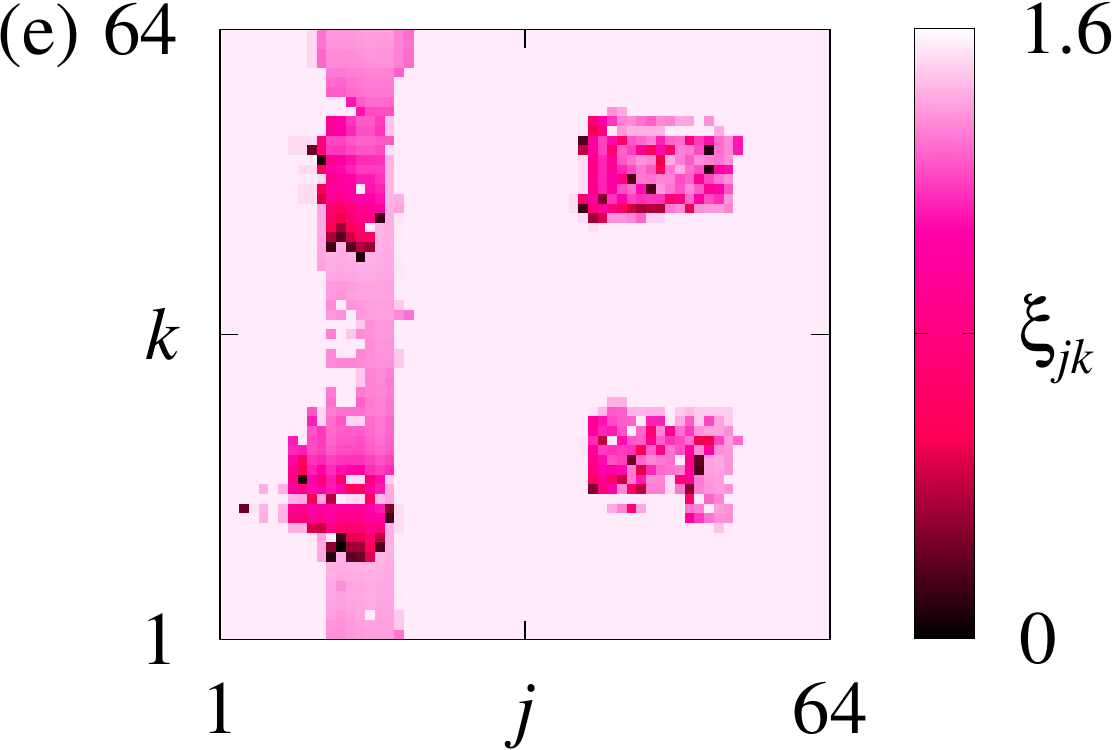}\hspace{2mm}
\includegraphics[height=0.21\textwidth]{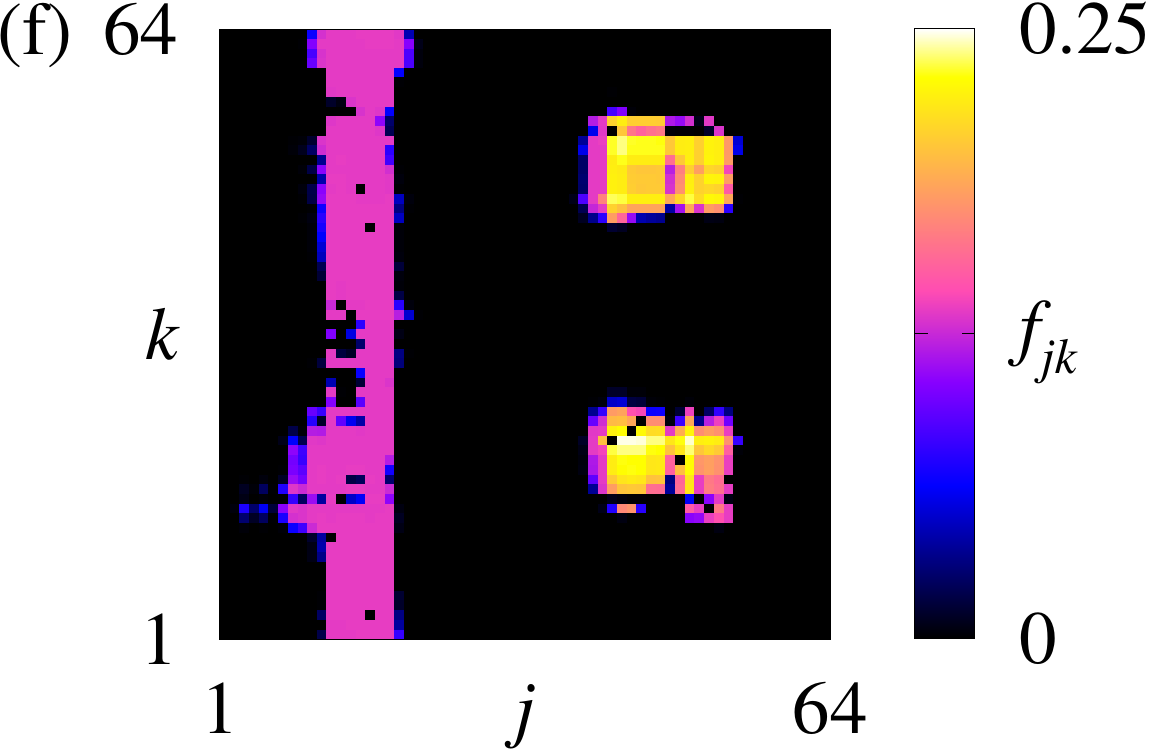}
\end{center}
\caption{\label{fig:d0_015}
(Color online)
Spatially localized (a-c) (see related video: Fig6\_movie.gif) and mixed bump states (d-f) in the LIF model~\eqref{eq03}:
(a,d) the potential $u_{jk}$, (b,e) the firing history function $\xi_{jk}$,
and (c,f) the firing rates $f_{jk}$ of the neurons.
All parameters are the same as in Fig.~\ref{fig:Tr0_0_}, 
except $d = 0.014$ (in panels (a-c)) and $d = 0.017$ (in panels (d-f)).
The time interval for calculation of $f_{jk}$ is $\Delta T = 1000$.
}
\end{figure}

\par As a first example, we consider two randomly generated realizations
of the uniform distribution of idle nodes for $d = 0.015$, see Fig.~\ref{fig:d0_015}.
Note that the actual number of idle elements (belonging to set $\mathbf{A}$)
equals $\mathbf{n}=58$ in panels (a-c) and $\mathbf{n}=71$ in panels (d-f),
which corresponds to 1.4\% ($d = 0.014$) and 1.7\% ($d = 0.017$), respectively,
of the total number of nodes $N\times N = 64\times 64 = 4096$.
Two dynamical regimes are observed in the resulting systems.
For each of these regimes we show its snapshot (a,d), firing history function (b,e),
and firing rates averaged over the time window $\Delta T = 1000$ (c,f).
All parameters are identical to the ones in Fig.~\ref{fig:Tr0_0_},
except for the addition of the above percentage $d$ of randomly introduced idle nodes
 which stay silent
at potential $u_{\rm idle}=u_{0}=0$ during the entire simulation.
Note that the position of the randomly selected idle nodes
in examples (a-c) and (d-f) is different.
It can be roughly recognized by the position
of the black points in snapshots (a) and (d),
because the active nodes take the reset value $u_0 = 0$
within a very short period of time due to their dynamics.

The snapshots in panels (a) and (d) of Fig.~\ref{fig:d0_015} look very similar:
Both of them appear as four-patch structures embedded in a, to a large extent,
homogeneous background.
However, these two states have qualitatively different collective dynamics.
This becomes clear if we look at the firing rates
shown in panels (c) and (f) of Fig.~\ref{fig:d0_015}.
Indeed, Fig.~\ref{fig:d0_015}(c) demonstrates
that the four active patches occupy almost fixed positions in space,
and therefore we observe a spatially localized state here.
In contrast, for the state shown in Fig.~\ref{fig:d0_015}(d-f),
only two patches have fixed positions in space,
while the other two move in the vertical direction, forming, thus, 
a ``mixed'' bump state, consisting of both localized and moving active domains.

\par
\begin{figure}[h]
\begin{center}
\includegraphics[height=0.21\textwidth]{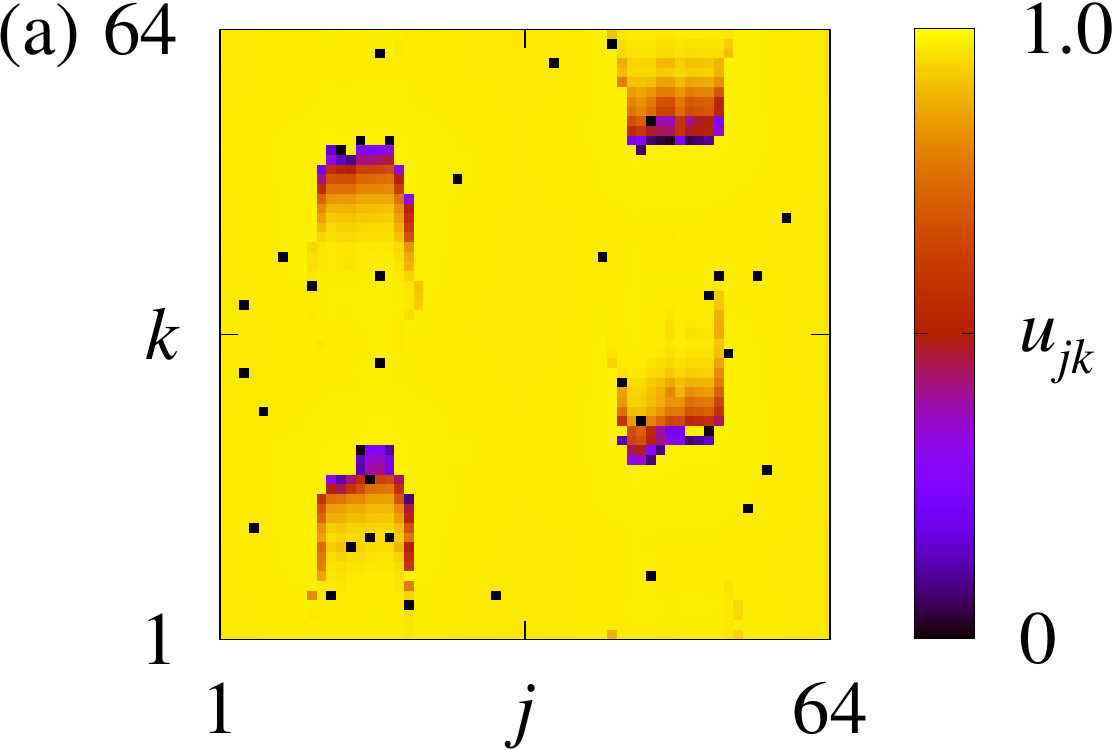}\hspace{2mm}
\includegraphics[height=0.21\textwidth]{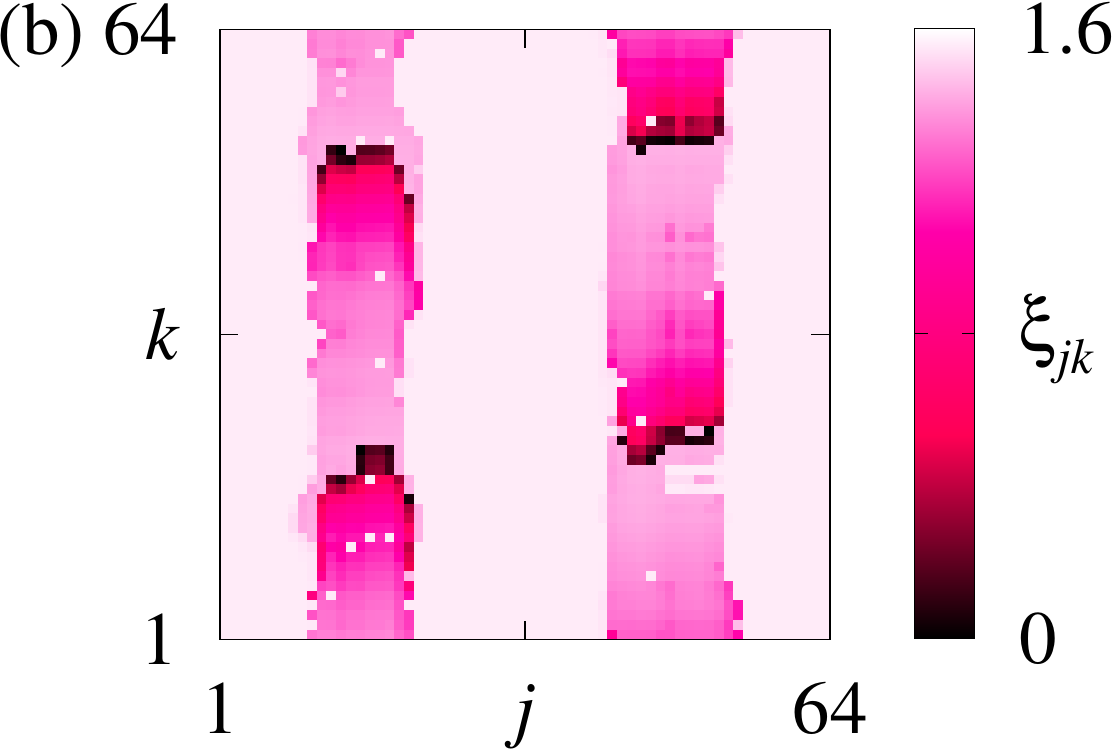}\hspace{2mm}
\includegraphics[height=0.21\textwidth]{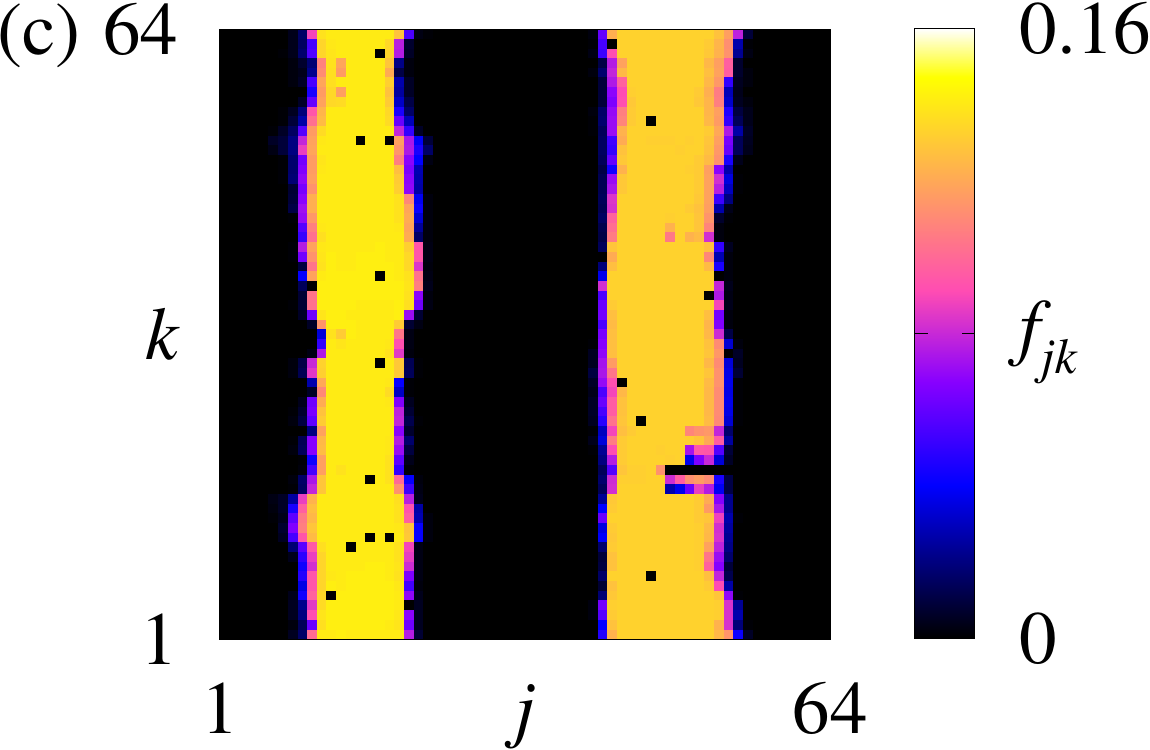}\\[2mm]
\includegraphics[height=0.21\textwidth]{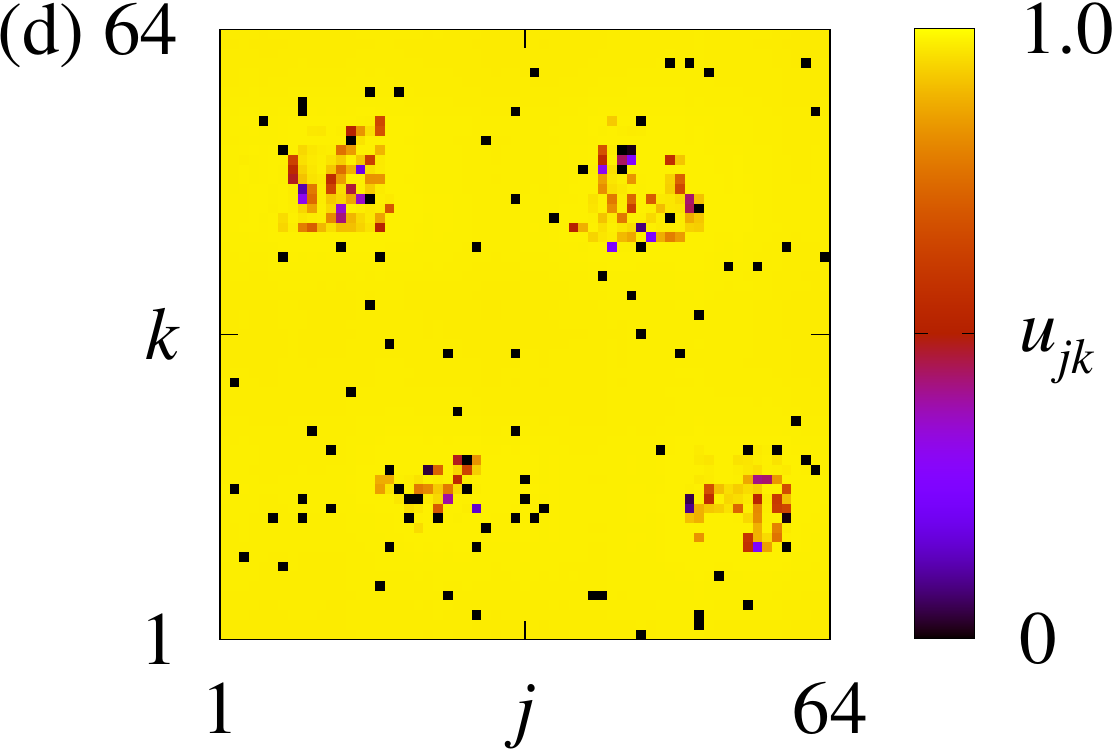}\hspace{2mm}
\includegraphics[height=0.21\textwidth]{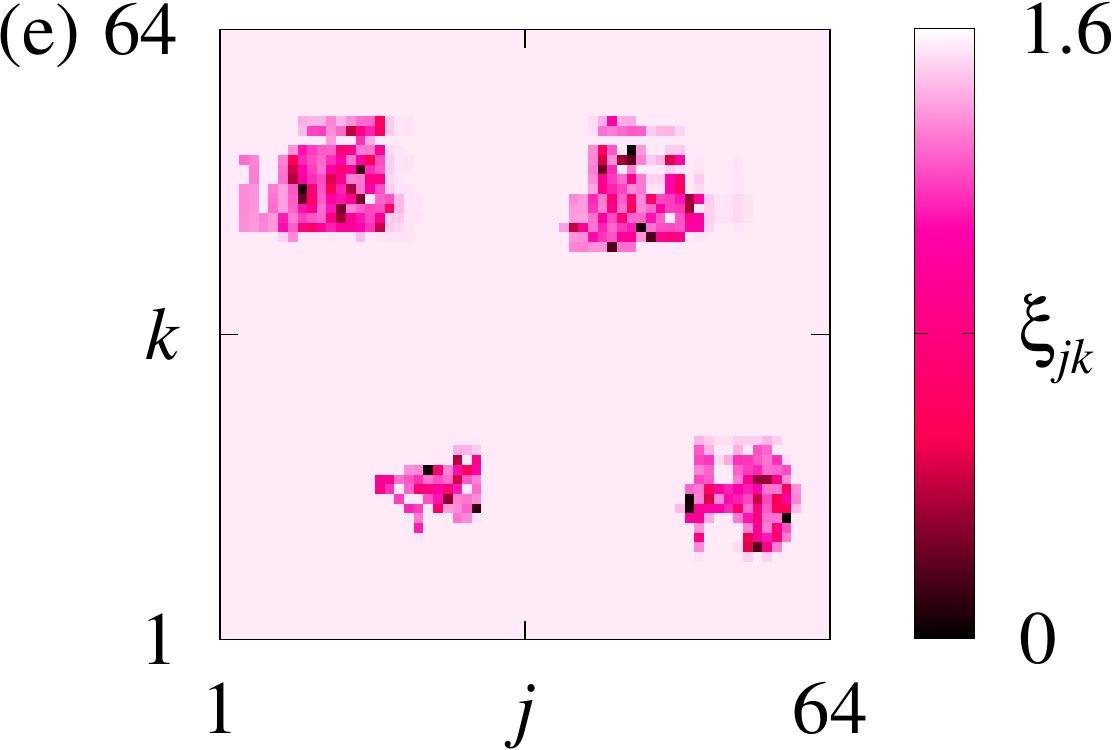}\hspace{2mm}
\includegraphics[height=0.21\textwidth]{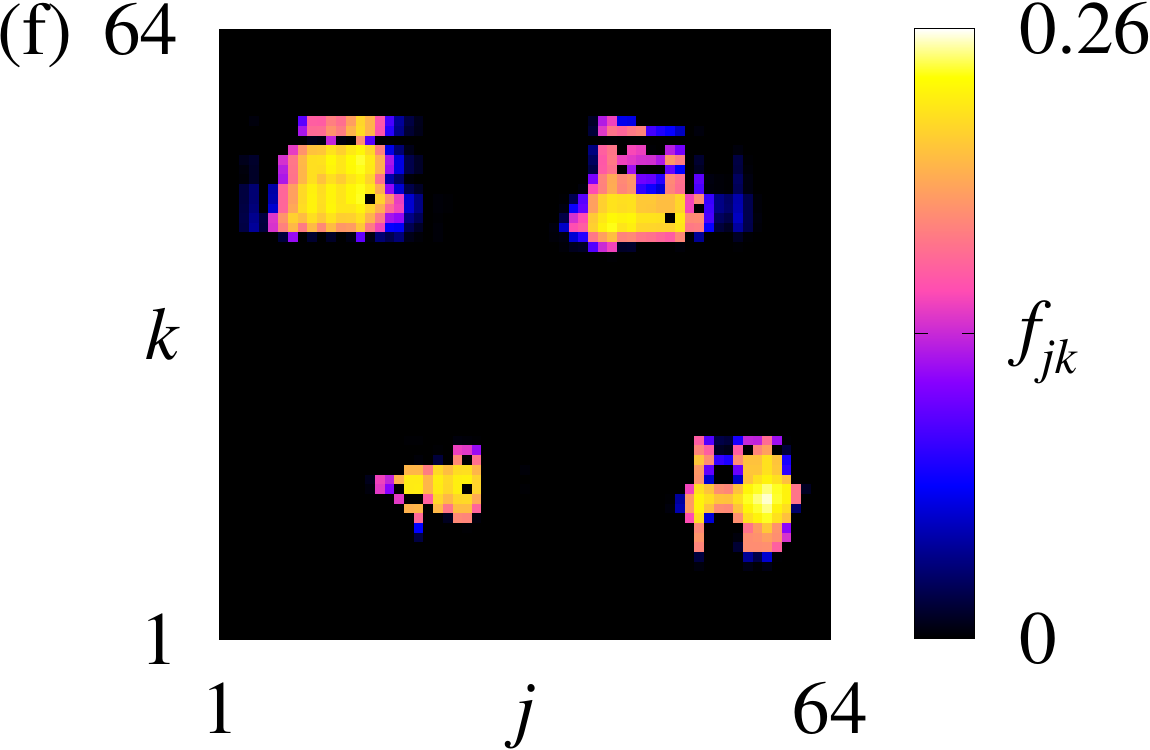}\\[2mm]
\includegraphics[height=0.21\textwidth]{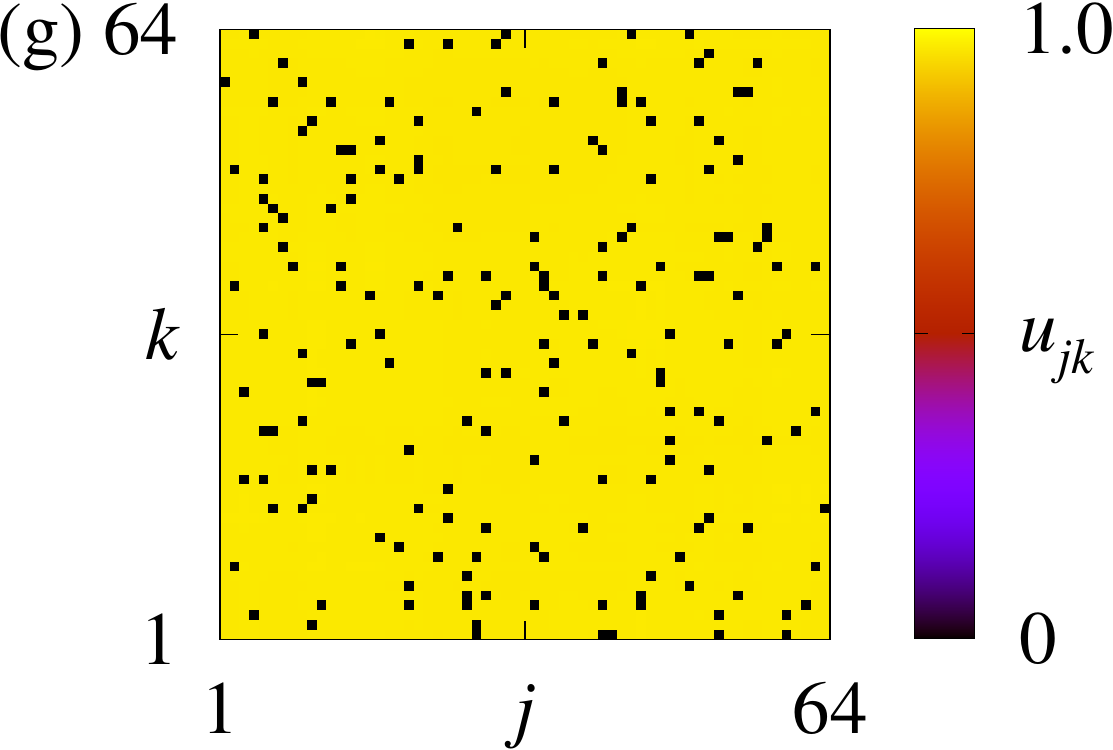}\hspace{2mm}
\includegraphics[height=0.21\textwidth]{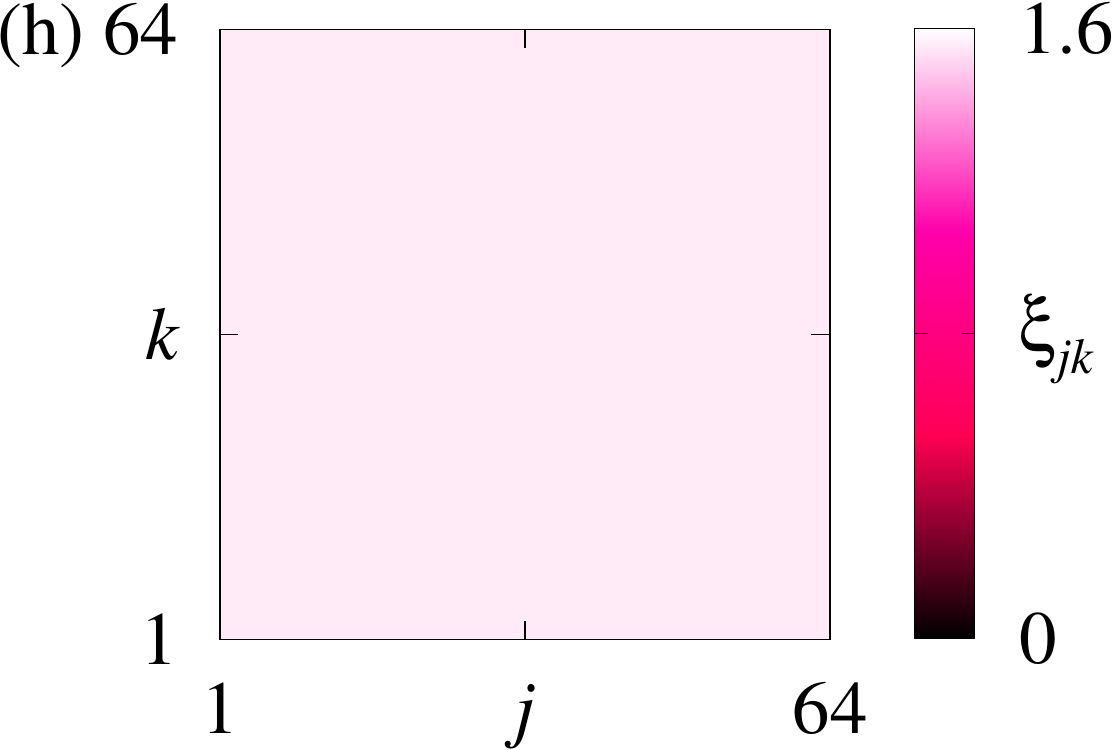}\hspace{2mm}
\includegraphics[height=0.21\textwidth]{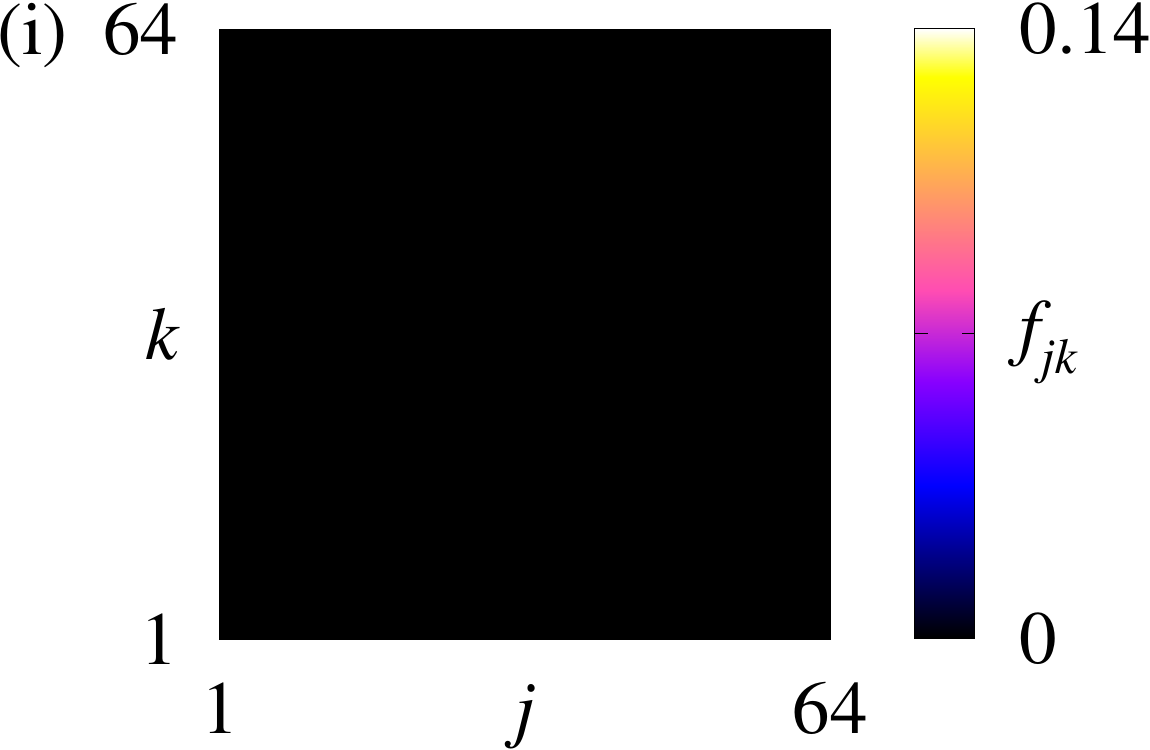}
\end{center}
\caption{\label{fig:d}
(Color online)
The effect of the density of idle nodes $d$ on the type
of predominant patterns observed in LIF model~\eqref{eq03}.
(a-c) Moving pattern for $d = 0.01$, (d-f) spatially localized pattern for $d = 0.02$,
and (g-i) static pattern for $d = 0.04$.
(a,d,g) the potential $u_{jk}$, (b,e,h) the firing history function $\xi_{jk}$,
and (c,f,i) the firing rates $f_{jk}$ of the neurons.
All parameters except $d$ are the same as in Fig.~\ref{fig:Tr0_0_}.
The time interval for calculation of $f_{jk}$ is $\Delta T = 1000$.
}
\end{figure}

 To investigate the influence of idle nodes in the network,
we present in Fig.~\ref{fig:d} three exemplary cases of low, medium and high
idle node concentration. In Fig.~\ref{fig:d}(a-c) we plot a typical state
of the network for low concentration of permanently idle nodes, $d=0.01$,
and we observe bumps moving in parallel to the y-direction. 
In the middle row of the same figure (panels (d-f)),
for intermediate $d$-values we observe localization of the bumps,
while in the bottom panels (g-i) we observe cessation 
of oscillations for large concentration of idle nodes.
It is interesting that when oscillations cease the node potentials attain fixed
(but not identical) values below the threshold. 
This phenomenon of full cessation
 of oscillations in the network is here 
recorded for the first time and is attributed to the presence of a 
relatively large number
(e.g., $d \ge 0.034$, for $R=12$) of permanently idle elements. 

\par To explore globally the $(R, d) $ parameter domains where bump localization takes place
in the case of idle nodes, 
numerical investigations were performed for a percentage of permanently idle nodes,  $d$, in the
range $0\le d \le 0.04$ and coupling ranges $1 \le R \le 15$.
To accelerate the simulations,
the system size used was $N \times N = 32 \times 32$ and in cases of doubt the behavior was 
confirmed using sizes $N \times N = 64 \times 64$.
The maximum coupling range
was limited to $R=15$ in the smaller, ($32 \times 32$), systems, which roughly corresponds to all-to-all connectivity, 
since $2R+1=31 \sim N=32$ (and similarly for the larger systems).
The maximum percentage of idle nodes was limited to $d=0.04$ because for these $d$ values
the numerical results showed that the entire system reaches a static state,
where all elements attain different fixed subthreshold values.

\par
\begin{figure}[h]
\includegraphics[clip,width=0.5\linewidth, angle=0]{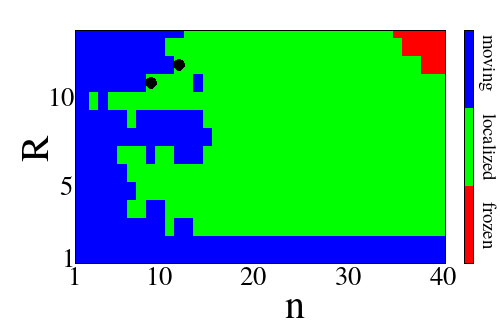}
\caption{\label{fig07} (Color online)
Color-coded map of moving or localized bump states in 
the LIF model in 2D with the presence of $\boldsymbol{n}\> (=dN^2)$ permanently idle nodes. 
Parameter regions where traveling bumps occur are depicted by blue color, 
localized bumps by green and cease of oscillations by red color.
The black dots denote the transitions from mobile to immobile bumps at
points ($R = 11, \boldsymbol{n}=9)$  and ($R = 12, \boldsymbol{n}=11)$, for comparison 
with later Fig.~\ref{fig08}(b).
Parameters are: $\mu=1.0$, $u_{\rm th}=0.98$, $u_0=0$, $\sigma=0.7$, $N\times N=32 \times 32$ and $T_\mathrm{r}=0$.
All simulations start from the same random initial conditions. 
}
\end{figure}

 Collectively, the results are
shown in Fig.~\ref{fig07}. For small values of $R$, e.g. $R= 1$, we are at the limit of
diffusion-like interactions and the network presents a large number of isolated 
active nodes moving around the network. A similar behavior is observed for $R=2$,
regardless of the density of idle nodes.
For $R\ge 3$ the active nodes cluster together to form domains/bumps. For small
numbers of randomly distributed idle nodes the bumps travel inside the
network. Such a directed motion was already seen in Fig.~\ref{fig:d}(a-c),
where the bumps on the left show a top-down motion,
while the ones on the right follow an opposite bottom-up route.
The number of idle elements is too small to be able to
intercept the motion of the bumps.
In the panel Fig.~\ref{fig:d}(a), we only added $41$ idle (black) nodes
and the bumps cross them straight through.
In the color-coded $(\boldsymbol{n},R)$ parameter plot, in Fig.~\ref{fig07},
the regions where the active domains are mobile are colored blue
and are located on the left and lower parts of the plot.
\par For intermediate values of $R$ and $\boldsymbol{n}$
the active domains get localized or confined around a central node;
the number of idle elements are now enough in numbers to intercept the bump motion.
The elements inside the active bumps oscillate
and they are surrounded by a sea composed of subthreshold and idle nodes, recall Fig.~\ref{fig:d}(d).
In the color-coded $(\boldsymbol{n},R)$ parameter plot (Fig.~\ref{fig07})
the regions where the active domains get localized are colored green.
As in the case of mechanism I (Refractory Period), Fig.~\ref{fig03}, 
the green color designates both the completely
localized, immobile bumps as well as the ones that move erratically but remain
confined near a certain position in the network.
For even larger numbers of idle elements and large connectivity ranges $R$,
oscillations do not survive in the system.
All elements (except for the idle ones) attain fixed subthreshold values
as can be seen in Fig.~\ref{fig:d}(g-i).
The parameter $(\boldsymbol{n}, R)$ region
where oscillations cease in the network is denoted with red color
on the right-top part in Fig.~\ref{fig07}.
\par
\begin{figure}[ht]
\includegraphics[clip,width=0.45\linewidth, angle=0]{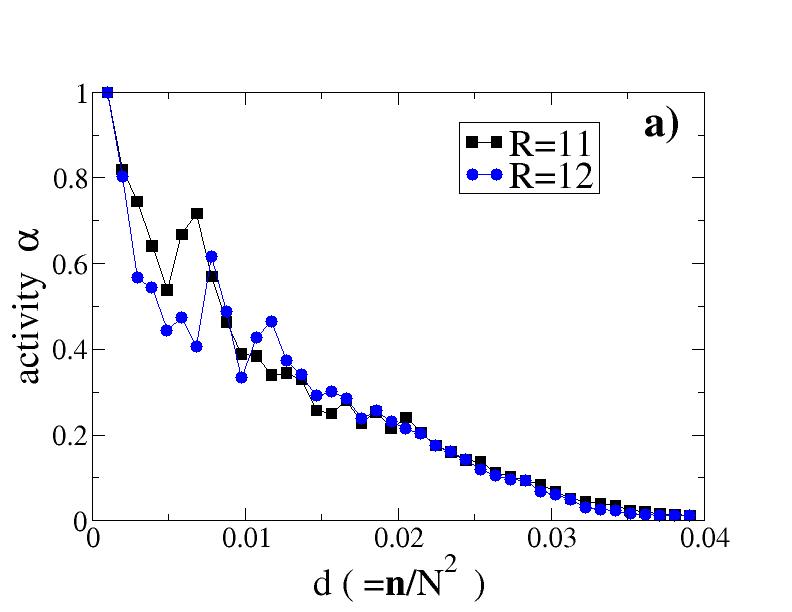}
\includegraphics[clip,width=0.45\linewidth, angle=0]{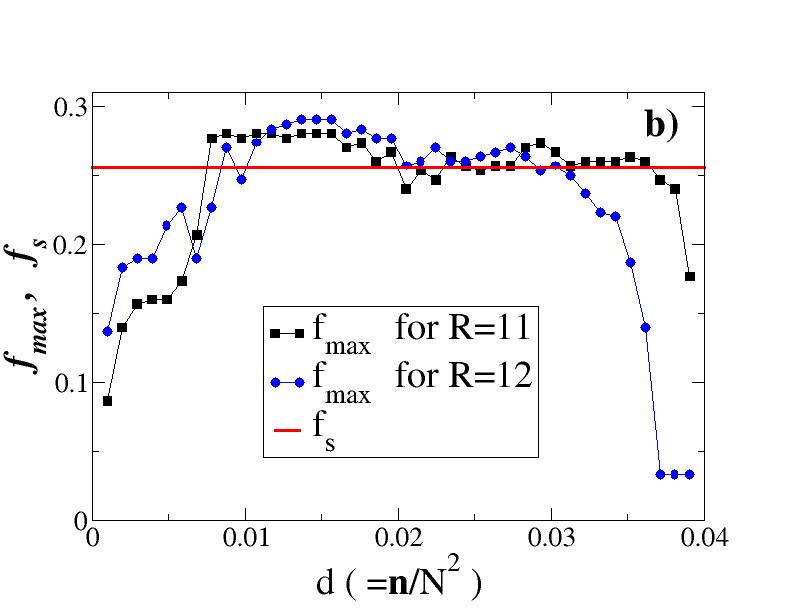}
\caption{\label{fig08} (Color online)
Moving/localized bump states in LIF model in 2D in the presence of $\boldsymbol{n}(=dN^2)$ permanently idle nodes. 
a) The activity $\alpha$ as a function of the density of the idle nodes, $d$, for $R=11$ (black squares)
and $R=12$ (blue circles) and b) The maximum firing rate $f_{\rm max}$ observed in the
network as a function of $d$, for $R=11$ (black squares)
and $R=12$ (blue circles). The solid red line in panel b) denotes the firing rate $f_\mathrm{s}$
of the uncoupled (single) LIF oscillator, without refractory period $T_\mathrm{r}=0$ 
($f_s$ value is calculated from Eqs.~\eqref{eq02} and ~\eqref{eq022}).
Other parameters are: $\mu=1.0$, $u_{\rm th}=0.98$, $u_0=0$, $\sigma=0.7$, $N\times N=32 \times 32$ and $T_\mathrm{r}=0$.
All simulations start from the same random initial conditions.}
\end{figure}

 To study the influence of the density of idle nodes on the activity of the network
we depict in Fig.~\ref{fig08}(a) the overall activity $\alpha =N_a/N^2$ of the system (see definition
in Sec.~\ref{sec:quantitative}) as a function of the density of idle nodes, $d$, for two
different values of the coupling range, $R=11,12$. As the plot indicates, the activity drops with
$d$, not only because of the presence of idle nodes but also because the idle nodes
influence their neighbors keeping them silent. This becomes most obvious especially in the 
presence of many dead nodes ($d\ge 0.036$), where the oscillations
cease in the entire network.
\par In Fig.~\ref{fig08}(b) we depict the maximum firing rates
in the network as a function of $d$. For small values of $d$ where traveling bumps are 
recorded (see Fig.~\ref{fig07}),  $f_{\rm max}$ takes small values 
because part of the time the nodes join the active domains
and the other part of the time they join the quiescent domain. 
This effect was also seen in the case of the refractory period, see Fig.~\ref{fig04}.  
In both cases (localization mechanisms I and II), 
the maximum firing rate values can differentiate between moving and localized
bumps in the network. As the number of idle nodes increase, $d>0.01$, and the bumps get
confined in space, the maximum firing rate $f_{\rm max}$ attains values as high as the
ones of the single LIF oscillator as may be seen from comparison with the level of $f_s$
(obtained via Eqs.~\eqref{eq02} and ~\eqref{eq022})
represented by the red solid line in Fig.~\ref{fig08}(b). For even higher $d$ values
the number and size of the bumps decreases (as also reflected by the decreasing activity in Fig.~\ref{fig08}(a)
leading to full cessation of oscillations. 

\par Overall, we may say that both mechanisms that cause localization take advantage of the presence of silent nodes
in the system. In mechanism I it is refractoriness which forces the nodes to spend a part of their time
in the rest state $u_{\rm rest}$=0 after resetting. In mechanism II some nodes, randomly chosen,
are constrained to be idle during their entire lifetime. Note that the two mechanisms originate from different
biological motivations: Mechanism I relates to refractoriness, a well known feature of all functional neurons, while
mechanism II relates to the presence of dead (idle) neuron cells.
 Nevertheless, in both cases by controlling the 
refractory period in case I or the number of idle nodes in case II, it is possible to control the propagation
of information (in the form of bumps/domains) 
and to modulate the overall system activity.

\section{Self-consistency equation for stationary bump states}
\label{sec:analytical}
All the above results regarding the dynamics of bump states
were obtained by numerical simulations.
Nevertheless, in the case of infinitely large networks
it is possible to develop a more
 theoretical approach to this problem.
We will show how this approach works by considering the simplest example ---
stationary bump states, i.e. localized activity patterns
that do not move relative to the network
and have statistically stationary properties over time.

\subsection{Derivation of the self-consistency equation}

Using the definition of the mean field potential~\eqref{Def:U},
we can rewrite Eq.~\eqref{eq03a} in the form
$$
\frac{du_{jk}}{dt} = \mu - u_{jk} + \sigma ( U_{jk} - u_{jk} ).
$$

\par
\begin{figure}[h]
\begin{tabular}{lclcl}
\includegraphics[height=0.2\textwidth]{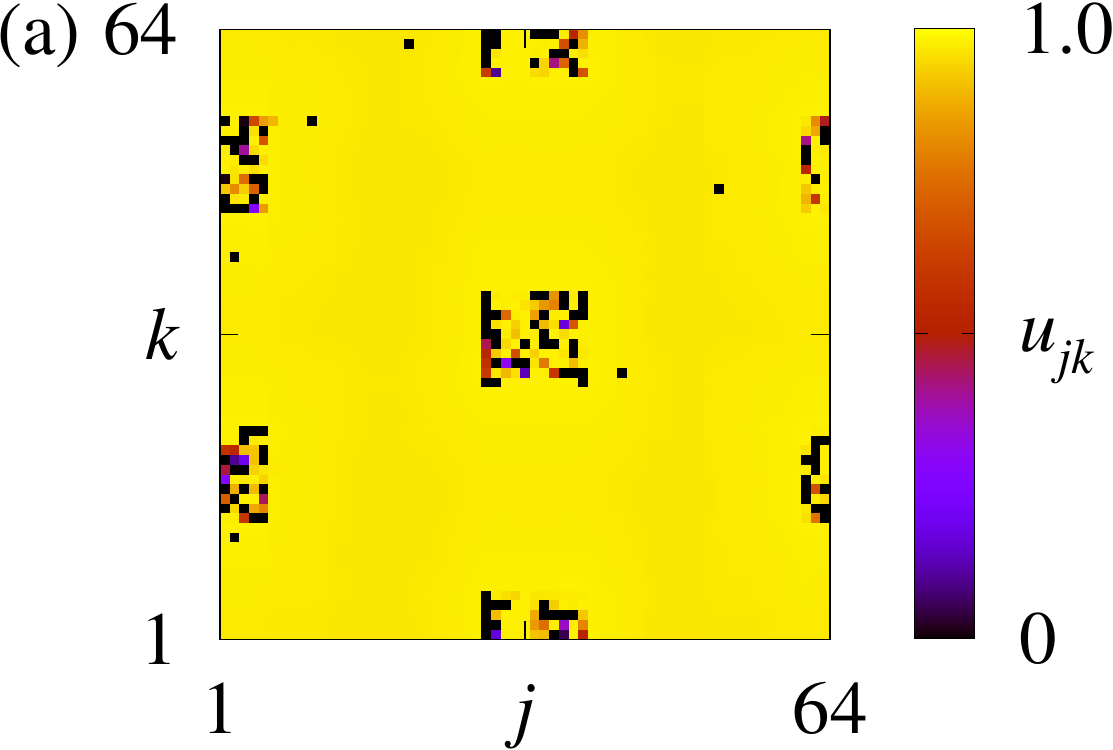} & &
\includegraphics[height=0.2\textwidth]{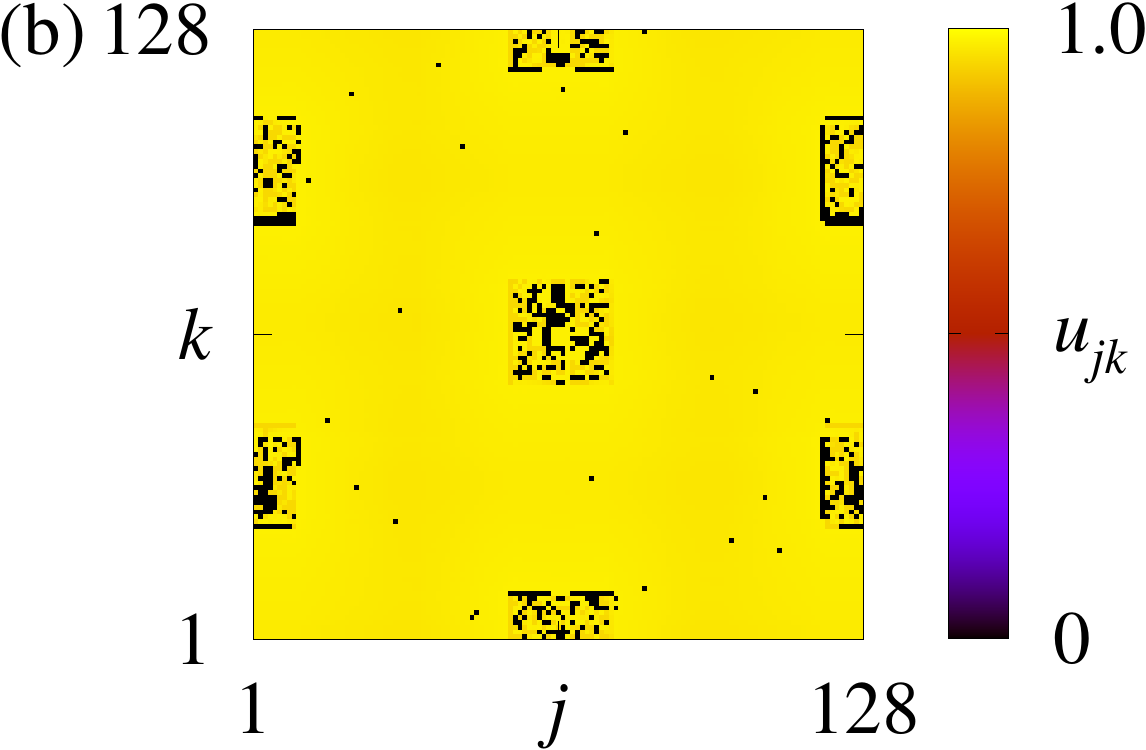} & &
\includegraphics[height=0.2\textwidth]{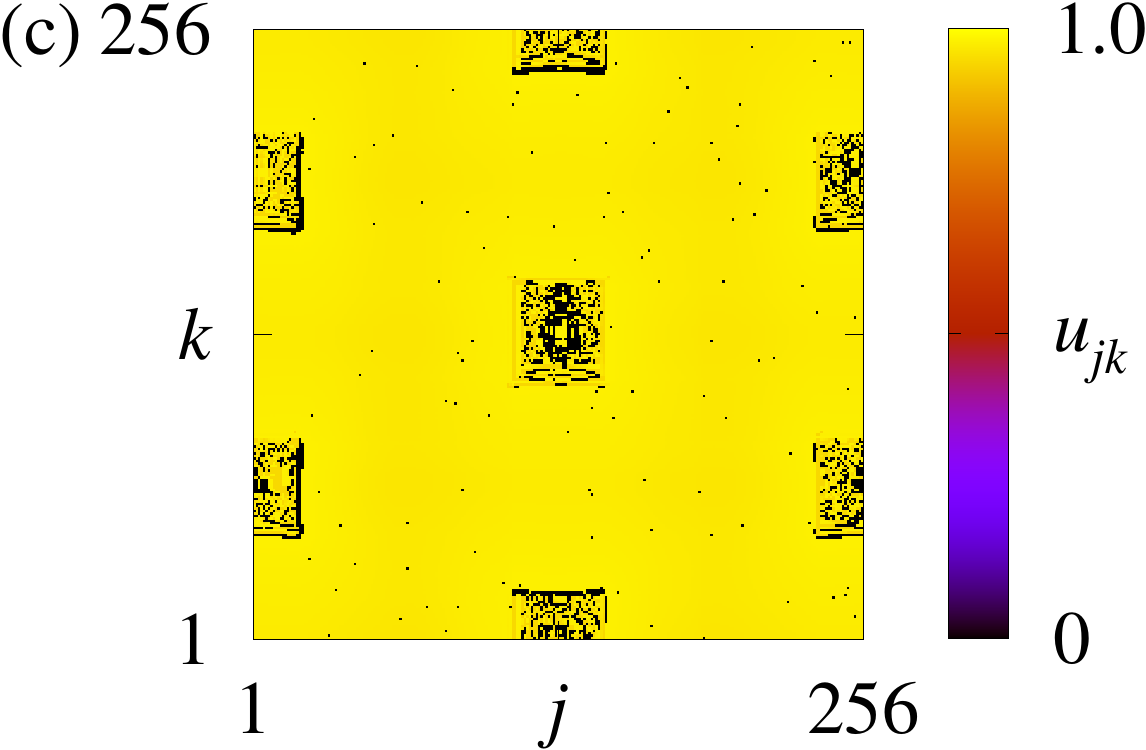}\\[2mm]
\includegraphics[height=0.2\textwidth]{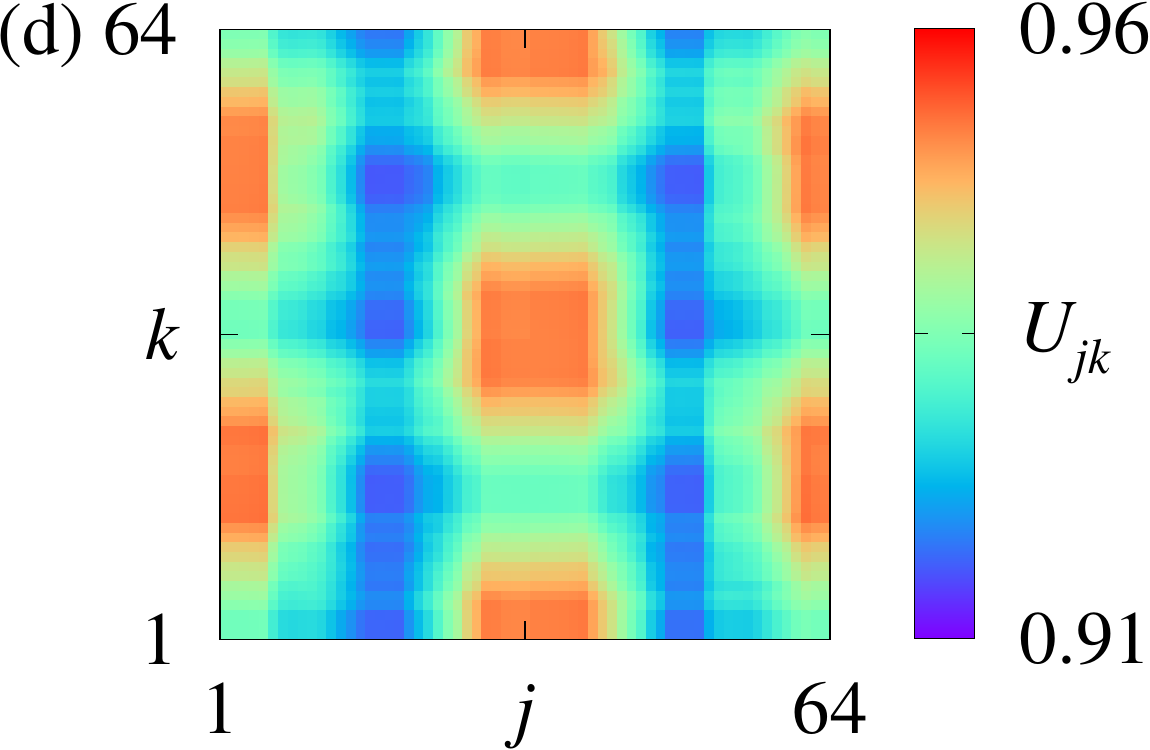} & &
\includegraphics[height=0.2\textwidth]{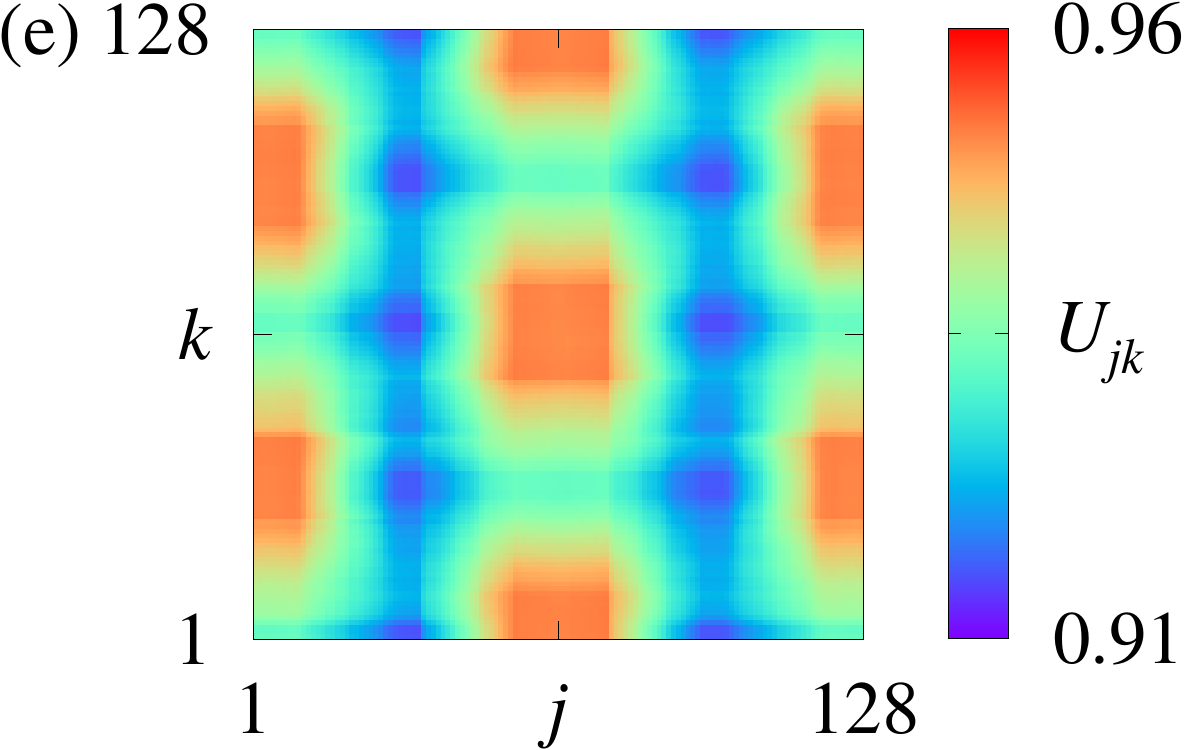} & &
\includegraphics[height=0.2\textwidth]{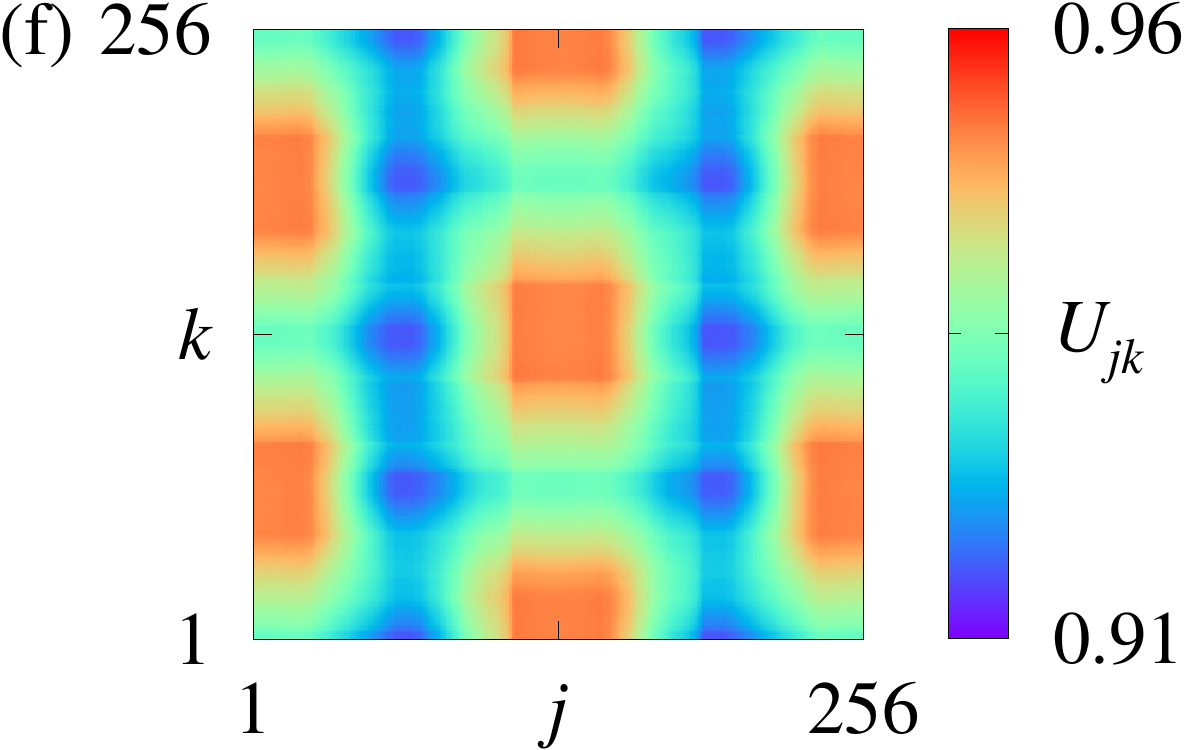}
\end{tabular}
\caption{\label{fig:Stationary}
(Color online) Scaling of the spatially localized patterns in the LIF model~\eqref{eq03}
for (a,d) $N = 64$, $R = 22$, (b,e) $N = 128$, $R = 44$, and (c,f) $N = 256$, $R = 88$.
Panels (a-c) show the potential $u_{jk}$
and panels (d-f) show the mean field potential $U_{jk}$ of the neurons.
All other parameters are the same as in Fig.~\ref{fig:Tr0_0_},
except $T_\mathrm{r} = 2.5$ and $d = 0.002$.
}
\end{figure}

 For some of the bump states found in system~\eqref{eq03}
we see that, after proper scaling,
their spatial profiles and their statistical properties
remain unchanged for any system size $N$,
provided that the ratio $R / N$ and other system parameters remain constant,
see Fig.~\ref{fig:Stationary}.
If in the large-$N$ limit, the mean field potential of such patterns
converges to a time-independent profile $U_\infty(x,y)$ such that
$$
\lim\limits_{N\to\infty} | U_{jk} - U_\infty(j/N,k/N) | = 0,
$$
then we call the corresponding pattern {\it stationary}.
For stationary patterns, it is possible to develop a self-consistency equation approach.
Roughly speaking, suppose that the scalar equation
\begin{equation}
\frac{du}{dt} = \mu - u + \sigma ( v - u )
\label{Eq:LIF:1D}
\end{equation}
with the reset condition~\eqref{eq03b} and the refractory period condition~\eqref{eq03c}
has a simple global attractor (i.e. an equilibrium or periodic orbit)
for any $v\in[u_0,u_\mathrm{th}]$.
Denoting the probability density of this attractor $\rho(u;v,\mu,\sigma,T_\mathrm{r})$,
we conclude that according to~\eqref{Def:U}
the stationary mean field potential profile $U_\infty(x,y)$ must satisfy
\begin{equation}
U_\infty(x,y) = d u_0 + \frac{1 - d}{(2\zeta)^2} \int\limits_{|x'-x|,|y'-y|\le\zeta} \int_{u_0}^{u_\mathrm{th}} u \rho(u;U_\infty(x',y'),\mu,\sigma,T_\mathrm{r}) du\: dx' dy'
\quad\mbox{for}\quad (x,y)\in[0,1]^2,
\label{Eq:SC}
\end{equation}
where $\zeta = R/N$ is the relative coupling length
and $d$ is the fraction of permanently idle elements.
In the next section, we will complete the description of Eq.~\eqref{Eq:SC}
by writing an explicit formula for $\rho(u;v,\mu,\sigma,T_\mathrm{r})$.

\subsection{Stationary dynamics of Eq.~\eqref{Eq:LIF:1D}}

Let us consider Eq.~\eqref{Eq:LIF:1D} with the parameters
$u_0 < u_\mathrm{th} < \mu$, $u_0\le v\le u_\mathrm{th}$, and $0 < \sigma < 1$.
Recall that the definition of this scalar equation
includes also the reset condition~\eqref{eq03b}
and the refractory period condition~\eqref{eq03c} with $T_\mathrm{r}\ge 0$.
It is easy to verify that for the above choice of parameters,
the dynamics of Eq.~\eqref{Eq:LIF:1D}
converges to one of two possible dynamical regimes:

(i) If $\mu + \sigma v < (1 + \sigma) u_\mathrm{th}$,
then Eq.~(\ref{Eq:LIF:1D}) has a stable equilibrium
$$
u_*(v,\mu,\sigma) = \frac{\mu + \sigma v}{1 + \sigma}.
$$
The corresponding stationary probability density is
$$
\rho(u;v,\mu,\sigma,T_\mathrm{r}) = \delta( u - u_*(v,\mu,\sigma) ).
$$

(ii) In the complementary case $\mu + \sigma v > (1 + \sigma) u_\mathrm{th}$,
the dynamics of Eq.~\eqref{Eq:LIF:1D} converges to a periodic orbit
that consists of active and refractory phases.
The duration of the active phase $T_\mathrm{a}$ depends
on the parameters $v$, $\mu$ and $\sigma$.
It can be calculated by analogy with Eq.~\eqref{eq02}, namely
$$
T_\mathrm{a} = \frac{1}{1+\sigma}  \ln\left( \frac{u_*(v,\mu,\sigma) - u_0}{u_*(v,\mu,\sigma) - u_\mathrm{th}} \right).
$$

The probability density of this phase reads
$$
\rho_\mathrm{a}(u;v,\mu,\sigma,T_\mathrm{r})
= \frac{C_\mathrm{a}}{\mu + \sigma v - (1 + \sigma) u}
= \frac{C_\mathrm{a}}{(1+\sigma)(u_*(v,\mu,\sigma) - u)}
$$
with the constant $C_\mathrm{a}$ determined by the normalization condition
$$
\int_{u_0}^{u_\mathrm{th}} \rho_\mathrm{a}(u;v,\mu,\sigma) du = \frac{T_\mathrm{a}}{T_\mathrm{a}+T_\mathrm{r}}.
$$
This yields

$$
C_\mathrm{a} = \frac{1}{T_\mathrm{a}+T_\mathrm{r}}.
$$

Taking into account that after each reset event the neuron stays at rest $u = u_0$
during the refractory period $T_\mathrm{r}$, we write the probability density
of the whole periodic trajectory as follows
$$
\rho(u;v,\mu,\sigma,T_\mathrm{r}) = \frac{T_\mathrm{r}}{T_\mathrm{a}+T_\mathrm{r}} \delta(u - u_0) + \frac{C_\mathrm{a}}{(1+\sigma)(u_*(v,\mu,\sigma) - u)}.
$$

Using these formulas for $\rho(u;v,\mu,\sigma,T_\mathrm{r})$, we calculate
$$
\int_{u_0}^{u_\mathrm{th}} u \rho(u;v,\mu,\sigma,T_\mathrm{r}) du = u_*(v,\mu,\sigma)
$$
if $\mu + \sigma v < (1 + \sigma) u_\mathrm{th}$.
Otherwise,

\begin{eqnarray*}
&&
\int_{u_0}^{u_\mathrm{th}} u \rho(u;v,\mu,\sigma,T_\mathrm{r}) du
= \frac{T_\mathrm{r}}{T_\mathrm{a}+T_\mathrm{r}} u_0
+ \int_{u_0}^{u_\mathrm{th}} \frac{C_\mathrm{a} u}{(1+\sigma)(u_*(v,\mu,\sigma) - u)} du\\[2mm]
&&
= \frac{T_\mathrm{r}}{T_\mathrm{a}+T_\mathrm{r}} u_0
- \frac{C_\mathrm{a}}{1+\sigma} \int_{u_0}^{u_\mathrm{th}} \frac{u}{u - u_*(v, \mu, \sigma)} du\\[2mm]
&&
= \frac{T_\mathrm{r}}{T_\mathrm{a}+T_\mathrm{r}} u_0
- \frac{C_\mathrm{a}}{1+\sigma} \left( u_\mathrm{th} - u_0
+ u_*(v, \mu, \sigma) \ln\left( \frac{u_\mathrm{th} - u_*(v, \mu, \sigma)}{u_0 - u_*(v, \mu, \sigma)} \right) \right)\\[2mm]
&&
= \frac{T_\mathrm{r}}{T_\mathrm{a}+T_\mathrm{r}} u_0
- \frac{T_\mathrm{a}}{T_\mathrm{a}+T_\mathrm{r}} \left( (u_\mathrm{th} - u_0)
\left[\ln\left(\frac{u_*(v,\mu,\sigma) - u_0}{u_*(v,\mu,\sigma) - u_\mathrm{th}}\right)\right]^{-1}
- u_*(v, \mu, \sigma) \right).
\end{eqnarray*}

The obtained results allow us to simplify the self-consistency equation~\eqref{Eq:SC}
by writing it in the form
\begin{equation}
U_\infty(x,y) = d u_0 + \frac{1 - d}{(2\zeta)^2} \int\limits_{|x'-x|,|y'-y|\le\zeta}
\mathcal{U}(U_\infty(x',y'),\mu,\sigma,T_\mathrm{r}) dx' dy'
\quad\mbox{for}\quad (x,y)\in[0,1]^2,
\label{Eq:SC_}
\end{equation}
where
$$
\mathcal{U}(v,\mu,\sigma,T_\mathrm{r}) = \left\{
\begin{array}{l}
u_*(v,\mu,\sigma),
\: \mbox{ if }\: \mu + \sigma v < (1 + \sigma) u_\mathrm{th}, \\[2mm]
 \frac{T_\mathrm{r}}{T_\mathrm{a}(v,\mu,\sigma)+T_\mathrm{r}} u_0
- \frac{T_\mathrm{a}(v,\mu,\sigma)}{T_\mathrm{a}(v,\mu,\sigma)+T_\mathrm{r}} \left( (u_\mathrm{th} - u_0)
\left[\ln\left(\frac{u_*(v,\mu,\sigma) - u_0}{u_*(v,\mu,\sigma) - u_\mathrm{th}}\right)\right]^{-1}
- u_*(v, \mu, \sigma) \right)\: \mbox{ otherwise}.
\end{array}
\right.
$$
Taking into account the fact that $U_\infty(x,y)$ is $1$-periodic with respect to $x$ and $y$,
we can solve Eq.~\eqref{Eq:SC_} approximately, using the Fourier-Galerkin method.
For the sake of simplicity, we explain how this method can be implemented
if $U_\infty(x,y)$ satisfies two additional symmetry conditions
$U_\infty(x,y) = U_\infty(-x,y) = U_\infty(x,-y)$. In this case, we can write
\begin{equation}
U_\infty(x,y) \approx \sum\limits_{n,m=0}^M \hat{u}_{nm} \phi_n(x) \phi_m(y)
\label{Ansatz:U}
\end{equation}
where $M\in\mathbb{N}$ and
$$
\phi_0(x) = 1,\quad \phi_n(x) = \cos(2\pi n x),\: n=1,2,\dots,
$$
are the elements of the $L^2$-orthogonal trigonometric basis, such that
$$
\int_0^1 \phi_n(x) \phi_{n'}(x) dx = \frac{\delta_{n n'}}{2 - \delta_{n 0}}.
$$

\begin{figure}[h]
\includegraphics[height=0.2\textwidth]{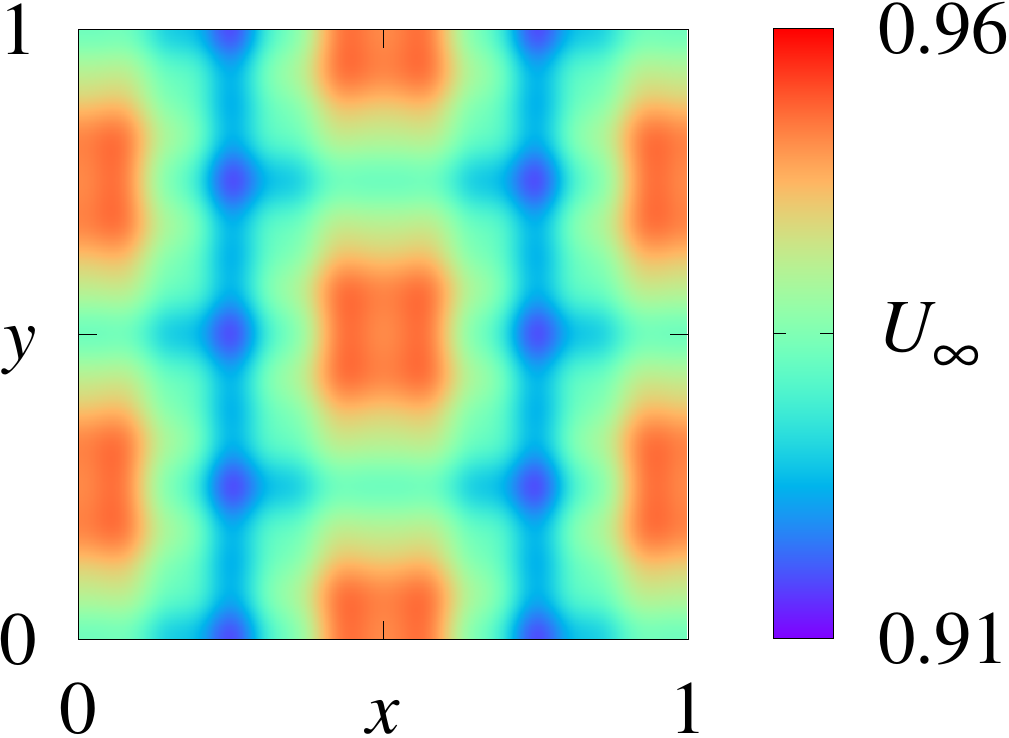}
\caption{\label{fig:MFP} (Color online)
Mean field potential $U_\infty(x,y)$ calculated
using Eq.~\eqref{System:SC} with $M = 10$ and ansatz~\eqref{Ansatz:U}.
Parameters: $\mu=1$, $u_\mathrm{th} = 0.98$, $u_0=0$, $\sigma=0.7$,
$\zeta = 11 / 32$, $T_\mathrm{r} = 2.5$ and $d=0.002$.}
\end{figure}

On the other hand, we can easily verify that the top-hat function
$$
H(x) = \left\{
\begin{array}{lcl}
1 / (2 \zeta) & \mbox{ if } & |x| \le \zeta,\\[2mm]
0 & \mbox{ if } &  \zeta < |x| \le 1/2
\end{array}
\right.
$$
(more precisely, its $1$-periodic extension from the interval $[-1/2,1/2]$)
can be written in the form of a Fourier series
$$
H(x) = \sum\limits_{n = 0}^\infty \hat{H}_n \phi_n(x)
$$
where
$$
\hat{H}_n = \int_{-1/2}^{1/2} H(x) \phi_n(x) dx = \left\{
\begin{array}{lcl}
1 & \mbox{ if } & n = 0,\\[2mm]
\frac{\sin(2 \pi n \zeta)}{\pi n \zeta} & \mbox{ if } &  n=1,2,\dots.
\end{array}
\right.
$$
Therefore, for every $1$-periodic function $u(x)$ we have
\begin{equation}
\int_0^1 \phi_n(x) \int_0^1 H(x - y) u(y) dy\: dx
= \frac{\hat{H}_n}{2 - \delta_{n 0}} \int_0^1 \phi_n(x) u(x) dx.
\label{Eq:Hn}
\end{equation}
Now, inserting ansatz~\eqref{Ansatz:U} into Eq.~\eqref{Eq:SC_},
projecting the resulting relation onto the basis function $\phi_n(x) \phi_m(y)$,
and using the identities~\eqref{Eq:Hn},
we obtain an approximating algebraic system
\begin{equation}
\hat{u}_{nm} = d u_0 \delta_{n 0} \delta_{m 0} + (1 - d) \hat{H}_n \hat{H}_m
\int_0^1 \int_0^1 \phi_n(x) \phi_m(y)\:
\mathcal{U}\left( \sum\limits_{n',m'=0}^M \hat{u}_{n' m'} \phi_{n'}(x) \phi_{m'}(y), \mu,\sigma,T_\mathrm{r}\right) dx\: dy
\label{System:SC}
\end{equation}
with $n,m = 0,1,\dots,M$.
To solve this system, we replaced the double integral in~(\ref{System:SC})
using Simpson's quadrature rule with $512$ points in the $x$- and $y$-directions,
and ran the standard Newton solver from MatLab
to find the unknown Fourier coefficients $\hat{u}_{nm}$.
Inserting these coefficients into~\eqref{Ansatz:U},
we obtained an approximate expression of the mean field potential $U_\infty(x,y)$.
For example, Fig.~\ref{fig:MFP} shows the mean field potential obtained
from Eq.~\eqref{System:SC} for the parameters corresponding to Fig.~\ref{fig:Stationary}.
The initial guess for the Newton solver was here obtained
by calculating the Fourier series of the snapshot in Fig.~\ref{fig:Stationary}(f).

\section{Conclusions and Open Problems }
\label{sec:conclusions}

In the present study the conditions for bump traveling and localization
were explored in 2D LIF networks of coupled neurons.
Two mechanisms of bump spatial confinement were proposed:
a) the introduction of a refractory period  (mechanism I) and
b) the inclusion of permanently idle nodes in the network (mechanism II). 
For both cases, we numerically determined the parameter domains where
bump localization or motion occurs. Overall, bumps tend to get localized in
systems with intermediate values of the coupling range $R$, high values of
refractory period $T_\mathrm{r}$ and high values of the density, $d$, of idle
nodes. In addition, for high values $R$ and $d$ the size of the active
domains shrinks and neural oscillations tend to cease throughout the system.
Note that although most of the above results were obtained for relatively small network sizes,
we also showed that in the large-$N$ limit,
the pattern formation in the LIF model~\eqref{eq03}
can be analyzed more rigorously using the self-consistency equation approach.

\par
We need to stress here that the directed motion of the asynchronous bumps
takes place even when the nodes have identical dynamics and are 
identically interacting, as in the cases presented above. If the network nodes
are not identical or the boundary conditions have special configurations,
 it is possible that the asynchronous oscillating
regions  will be localized. 
It remains an open problem to identify special boundary conditions 
or connectivity schemes
which give rise to localized bump states without the need for a refractory period
or for the introduction of random idle elements.

\par
\begin{figure}[h]
\begin{center}
\includegraphics[height=0.21\textwidth]{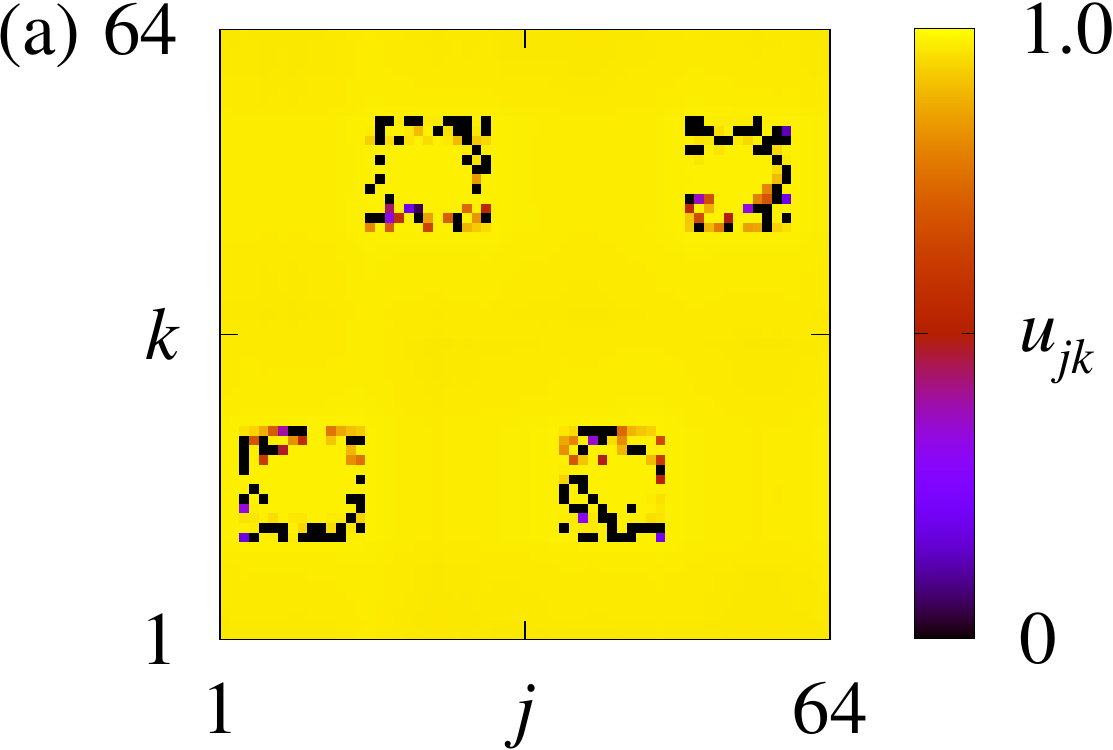}\hspace{2mm}
\includegraphics[height=0.21\textwidth]{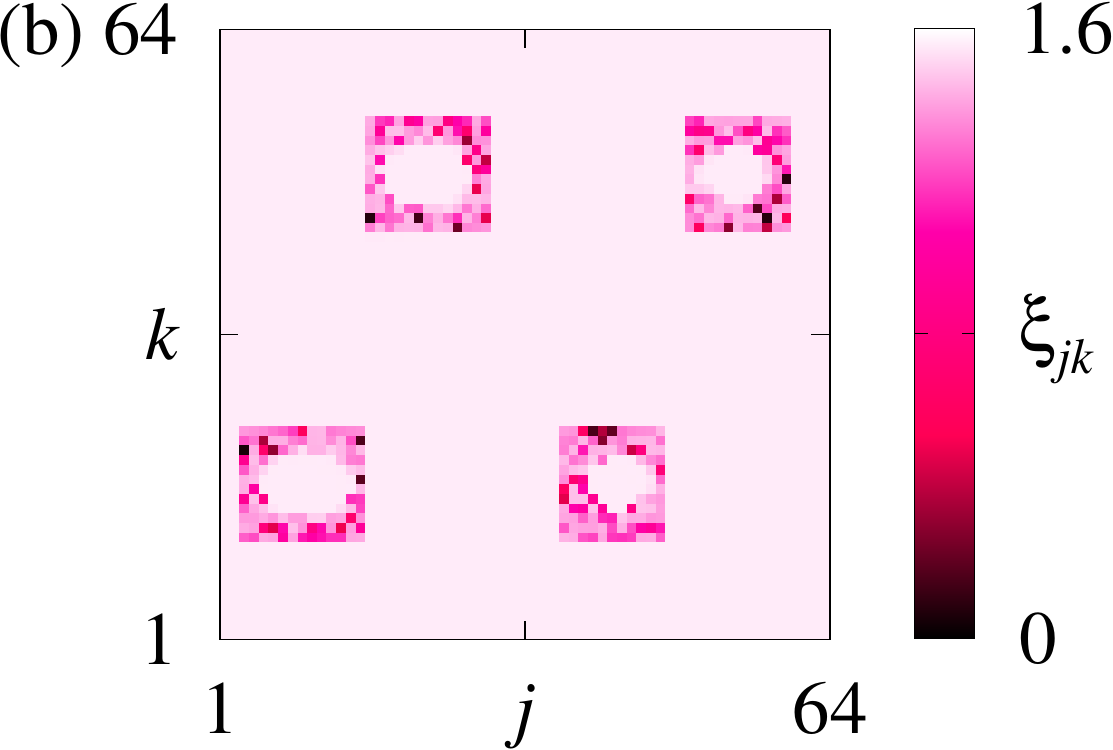}\hspace{2mm}
\includegraphics[height=0.21\textwidth]{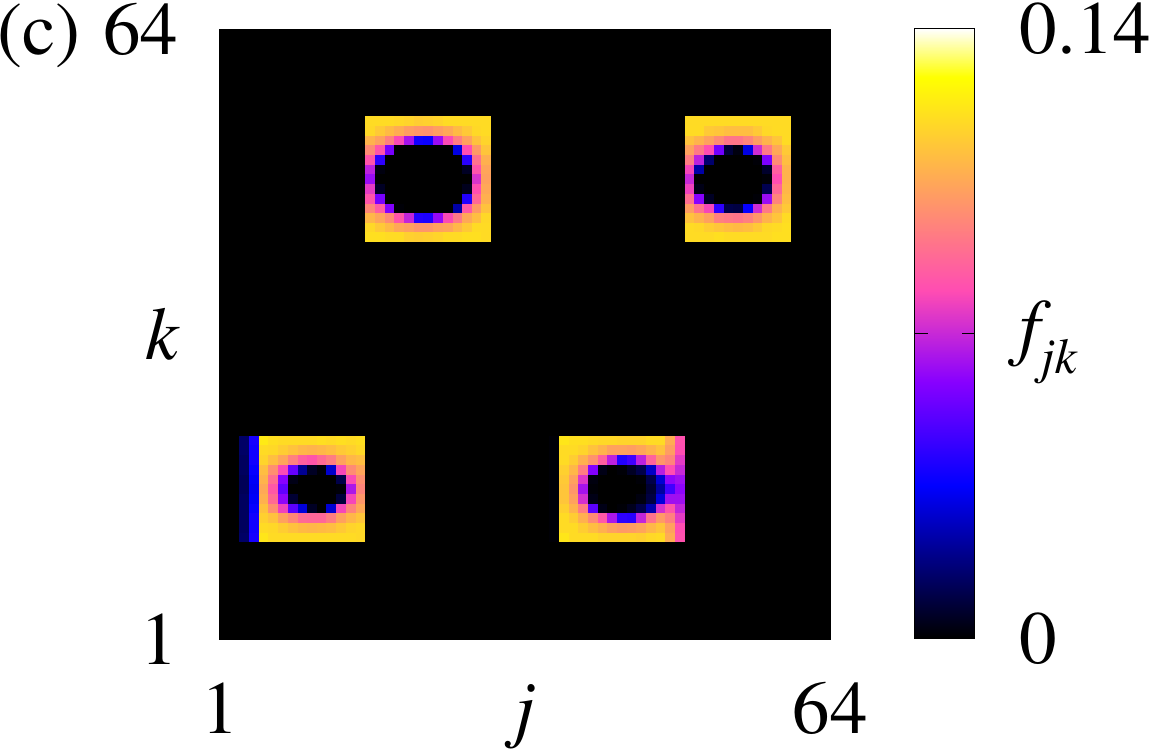}
\end{center}
\caption{\label{fig:Tr3_0R20}
(Color online)
Spatially localized hollow  bump states in the LIF model~\eqref{eq03}:
(a) the potential $u_{jk}$, (b) the firing history function $\xi_{jk}$,
and (c) the firing rates $f_{jk}$ of the neurons. See related video fig12\_movie.gif.
All parameters are the same as in Fig.~\ref{fig:Tr0_0_},
except for $R = 20$ and $T_\mathrm{r} = 3$. 
The time interval for calculation of $f_{jk}$ is $\Delta T = 1000$.
}
\end{figure}

 It is important to note that the bump states in model~\eqref{eq03}
can have a more complex structure than in Fig.~\ref{fig:Tr2_5}.
For example, in the case $N\times N = 64\times 64$, $R = 20$,
and $T_\mathrm{r} = 3$, we found a ring bump state,
where each of the four active regions has a ring structure
with a quiescent spot in the center of the bump,
see Fig.~\ref{fig:Tr3_0R20}.

\par
\begin{figure}[h]
\begin{center}
\includegraphics[height=0.21\textwidth]{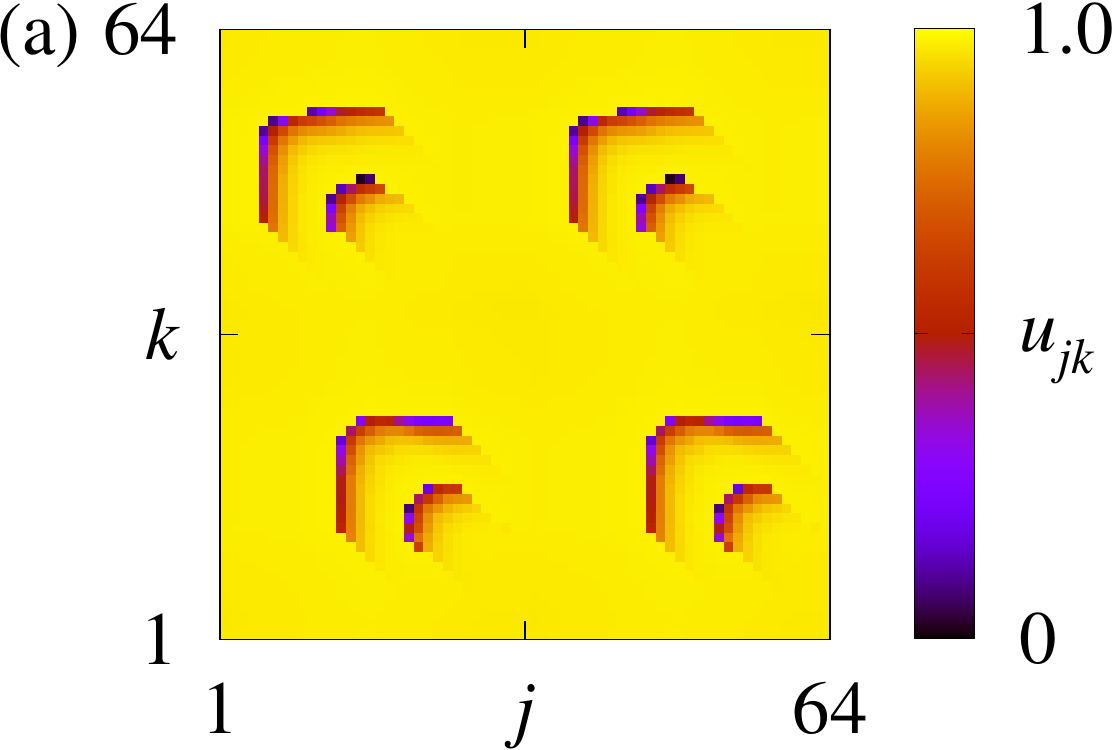}\hspace{2mm}
\includegraphics[height=0.21\textwidth]{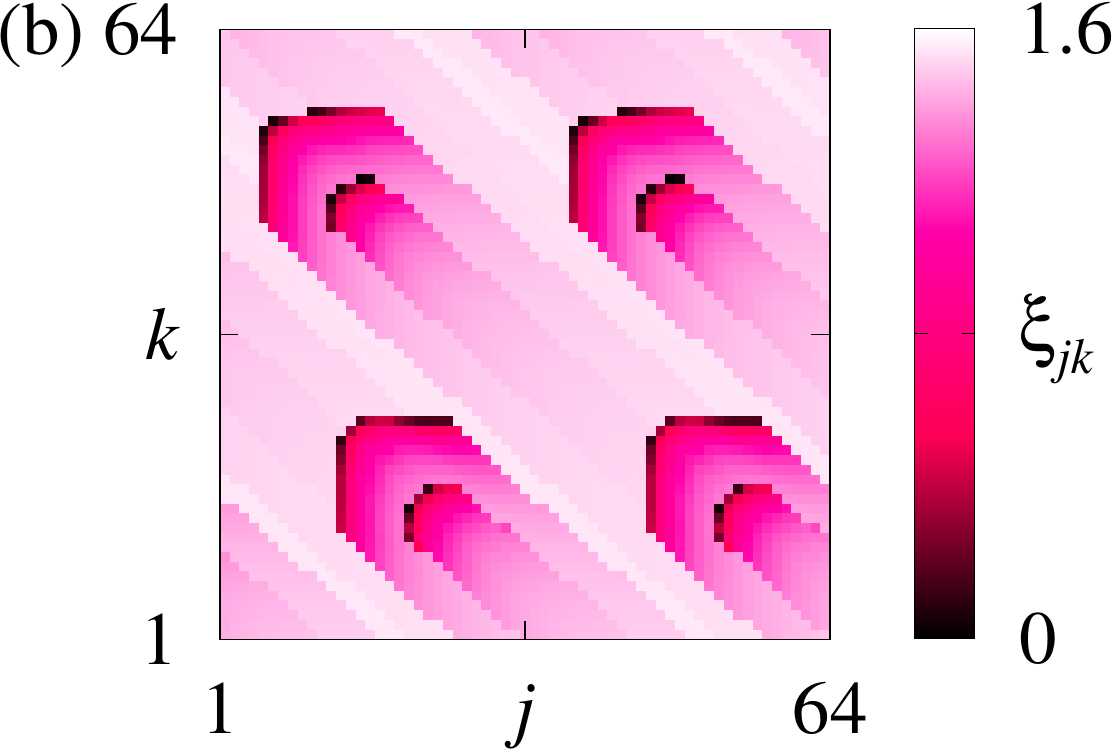}
\end{center}
\caption{\label{fig:Tr0_0R20}
(Color online)
Traveling double-arrow pattern in the LIF model~\eqref{eq03}:
(a) the potential $u_{jk}$, and (b) the firing history function $\xi_{jk}$ of the neurons.
See related video fig13\_movie.gif.
All parameters are the same as in Fig.~\ref{fig:Tr0_0_}, but with $T_\mathrm{r} = 0$ and $R = 20$.
}
\end{figure}

 Remarkably, in this case we again see the critical role of the refractory time
in the stabilization of bump states.
Indeed, if we run the same simulation, but with $T_\mathrm{r} = 0$,
we observe the formation of traveling bump states shown in Fig.~\ref{fig:Tr0_0R20}
(traveling double-arrows).

\par
It would be interesting to study how the bump states and bump localization is 
achieved in other neural models, where the
refractory period is not added via a condition as in the LIF model, 
but it arises due to the nonlinearity 
of the model equations. Indicatively, such models are the FitzHugh-Nagumo, the Hindmarsh-Rose and 
the Hodgkin-Huxley neural oscillators.
For these models the duration of refractoriness is controlled by specific system parameters. 
It is an open problem
to explore whether the localization of bumps in these alternative neural models 
occurs within parameter ranges that produce extended 
refractory periods and to establish a possible analogy between neural refractoriness 
(a property of the individual neurons)  and bump localization (a collective property of the network).

\par In relation to mechanism II (presence of idle nodes), it would be interesting 
to address the case where the link weights decay, approaching zero. 
This is an altogether different scenario
because a connection $\sigma_{mn;jk}$ of strength $0$
between nodes $(j,k)$ and $(m,n)$, penalizes only 
the single difference $\left[ u_{mn}-u_{jk}\right]$ in the interaction terms
of Eq.~\eqref{eq03a}, whereas if one neuron/node $(j,k)$ is set to the ``dead'' state, 
then this affects all the connections formed by this neuron.
 This connection-decay scenario
may lead to bump states with entirely different mobility properties than the ones
described in Sec.~\ref{sec:idle}.

 \par In the presence of random idle nodes (mechanism II), it would also be interesting
to study the collective properties of the system near the percolation threshold,
where the idle nodes form a large connected cluster which extends over the entire system.
Nevertheless, in the present study the cessation 
 of oscillations seems to take place
far below the percolation threshold. However, the network that was used here was
of finite size. Larger networks need to be considered not only to address the question
on the percolation threshold but also to study the finite size effects in the
propagation of the bumps in the system.

\section{Acknowledgments}
This work was 
supported by computational time granted from the Greek Research \& Technology Network (GRNET)
in the National High Performance Computing HPC facility - ARIS - under project ID PR014004.
The work of O.E.O. was supported by the Deutsche Forschungsgemeinschaft under Grant
No. OM 99/2-3.

\section*{Data Availability Statement}
Data available on request from the authors.
%
%

%

\end{document}